\documentclass[11pt, onecolumn, article]{ieeetran}
\usepackage{cite}
\usepackage[cmex10,intlimits]{amsmath}
\usepackage{amsfonts}
\usepackage{graphicx, color, psfrag}
\usepackage{epsfig}
\usepackage{amssymb}
\usepackage{amsthm}
\usepackage[body={6.5in,9in},top=1in, left=1in]{geometry}
\usepackage{mathrsfs}

\newcommand{\vect}[3]{[\begin{array}{ccc} #1 & #2 & #3 \end{array}]}
\newcommand{\letca}{C_{0}} 
\newcommand{\letcb}{C_{1}} 
\newcommand{\ththreeconst}{C_3}
\newcommand{\thfourconst}{C_4}
\newcommand{\sdtheoremcb}{C_6}
\newcommand{\letcpref}{C_{7}} 
\newcommand{\letcf}{C_{8}} 
\newcommand{\letcg}{C_{9}} 
\newcommand{\letch}{C_{10}} 
\newcommand{\letcd}{C_{11}} 
\newcommand{\sdtheoremca}{C_{12}} 

\newcommand{\letpa}{\Psi_0} 
\newcommand{\letpththree}{\Psi_3} 
\newcommand{\sdtheorempb}{\Psi_{6}} 

\newcommand{\letpb}{\Psi_{7}} 
\newcommand{\letpc}{\Psi_{8}} 
\newcommand{\letpd}{\Psi_{9}} 
\newcommand{\letpe}{\Psi_{10}} 
\newcommand{\letpf}{\Psi_{11}} 
\newcommand{\letpg}{\Psi_{12}} 
\newcommand{\sdtheorempa}{\Psi_{13}} 
\newcommand{\sdtheorempc}{\Psi_{14}}

\newcommand{\mtb}{\mathcal{T}_2}
\newcommand{\mtc}{\mathcal{T}_4}
\newcommand{\mtd}{\mathcal{T}_5}
\newcommand{\mtf}{\mathcal{T}_6}

\newtheorem{theorem}{Theorem}
\newtheorem{proposition}{Proposition}
\newtheorem{lemma}{Lemma}

\newtheorem{definition}{Definition}
\newtheorem{conjecture}{Conjecture}
\newtheorem{example}{Example}
\newcommand{\diversity}{\mathtt{d}}

\title{The Necessity of Relay Selection}
\begin{document}
\author{Erdem Koyuncu, and Hamid Jafarkhani, \emph{Fellow}, IEEE\thanks{The authors are with the Center for Pervasive Communications and Computing, University of California, Irvine, Irvine CA 92697-2625 USA. Email: \{ekoyuncu, hamidj\}@uci.edu.}}

\maketitle
\begin{abstract}
%
%
We determine necessary conditions on the structure of symbol error rate (SER) optimal quantizers for limited feedback beamforming in wireless networks with one transmitter-receiver pair and $R$ parallel amplify-and-forward relays. We call a quantizer codebook ``small'' if its cardinality is less than $R$, and ``large'' otherwise. A ``d-codebook'' depends on the power constraints and can be optimized accordingly, while an ``i-codebook'' remains fixed. It was previously shown that any i-codebook that contains the single-relay selection (SRS) codebook achieves the full-diversity order, $R$. We prove the following:

Every full-diversity i-codebook contains the SRS codebook, and thus is necessarily large. In general, as the power constraints grow to infinity, the limit of an optimal large d-codebook contains an SRS codebook, provided that it exists. For small codebooks, the maximal diversity is equal to the codebook cardinality. Every diversity-optimal small i-codebook is an orthogonal multiple-relay selection (OMRS) codebook. Moreover, the limit of an optimal small d-codebook is an OMRS codebook.

We observe that SRS is nothing but a special case of OMRS for codebooks with cardinality equal to $R$. As a result, we call OMRS as ``the universal necessary condition'' for codebook optimality. Finally, we confirm our analytical findings through simulations.
\end{abstract}
\section{Introduction}
\IEEEPARstart{T}{he} availability of channel state information (CSI) can greatly affect the performance and reliability of amplify-and-forward (AF) cooperative relay networks. With available CSI, each relay can adaptively adjust its transmit power and transmit phase. This network beamforming scheme has been shown to achieve maximal diversity and array gains\cite{larsson1, jing1, koyuncu1}. In contrast, without any CSI at the relays, distributed space-time coding schemes can also achieve maximal diversity, but they also incur an unbounded array gain loss compared to network beamforming\cite{laneman2, jing4}.

For networks with parallel relays, the optimal beamforming policy requires one or two real numbers to be broadcasted from the receiver to the relays. A more practical assumption is that there is only partial CSI at the relays. For such networks, it has been shown that beamforming with quantized instantaneous CSI can achieve not only the maximal diversity gain but also a very high array gain with only a few feedback bits\cite{koyuncu1, zhao5}.

A special case of quantized feedback for relay networks is single-relay selection (SRS) \cite{bletsas1, riberio1,anghel1,hasna1,zhao3, zhao6, jing6}, which uses $\lceil \log_2 R \rceil$ feedback bits per channel state for a network with $R$ relays. It allows only one of the relays to cooperate given a constant fading block. This simple quantization scheme has been shown to achieve full-diversity for a very broad class of network topologies\cite{koyuncu1, shi1, koyuncu3}, and even under suboptimal selection criteria\cite{jing6}.

In this work, we consider a network with one transmitter-receiver pair and $R$ parallel AF relays. We assume that there is no direct link between the transmitter and the receiver. The transmitter and the relays have their own short-term power constraints. We assume that the receiver has full CSI, while each relay knows only the magnitude of its own receiving channel and has $B$ bits of partial CSI. The feedback bits are conveyed from the receiver to the relays via error-free and delay-free feedback channels, and they represent a quantized beamforming vector. Our performance measure is the symbol error rate (SER).

A well-known performance measure that is closely related to SER is diversity. We define the diversity measure for our network model as follows: Let $P_0$ and $P_i,\,i=1,\ldots,R$ represent the transmitter and relay power constraints, respectively. We allow these power constraints to vary linearly with a common power constraint $P$ as $P_i \triangleq p_i P,\,i=0,\ldots,R$, where $p_i$ are fixed positive real numbers that are independent of $P$. Then, as $P\rightarrow\infty$, the SER converges to $\mathtt{a}P^{-\mathtt{d}}$, where $\mathtt{a}$ and $\mathtt{d}$ represent the \textit{array gain}, and the \textit{diversity gain}, respectively. Since there are $R$ independently fading paths between the transmitter and the receiver, the maximal spatial diversity of our network is $R$, which we call the full-diversity order.

The set of all $2^B$ quantized beamforming vectors is the quantizer codebook. For clarity of exposition, we classify the codebooks under two criteria, one of which is cardinality: We call a codebook ``small'' if its cardinality is less than the number of relays, and ``large'' otherwise. We shall see later on that it is necessary to use a large codebook in order to achieve full-diversity, and correspondingly, the diversity provided by a small codebook is strictly less than $R$.

The other criterion that we use is the codebooks' dependence on the transmitter and relay power constraints, the motivation of which we now explain. In general, we can optimize the codebook with respect to the power constraints, as demonstrated in \cite{koyuncu1}. We call such power-dependent codebooks as ``d-codebooks''. Note that, an optimal codebook given some power constraints will lose its optimality as soon as any of the constraints are changed. Then, in order to achieve the best performance for any choice of constraints, the receiver and the relays need to store a possibly large number of optimal codebooks. A more practical approach might be to consider a power-independent codebook (i-codebook). In this case, a single codebook is used for all possible constraints with the purpose of achieving high diversity and array gains.

%
%
%
%

The main contributions of this paper can be summarized as follows: First, we show that every full-diversity i-codebook necessarily contains the SRS codebook. We obtain an analogous result for power-dependent codebooks: As $P\rightarrow\infty$, the limit of an optimal large d-codebook contains an SRS codebook, provided that it exists. Both results show that full-diversity codebooks should incorporate the SRS codebook structure, and are necessarily large.

For small codebooks, we show that the maximal achievable diversity is equal to the cardinality of the codebook. We would like to note that, even though this result is well-known for the case of limited feedback beamforming in multiple-input single-output (MISO) systems\cite{mukkavilli1,roh1, love1, koyuncu2}, its proof requires a completely different approach in our case.

Having determined the best achievable diversity of small codebooks, we show that any optimal small i-codebook is an ``orthogonal multiple-relay selection'' (OMRS) codebook,
meaning that it consists of multiple-relay selection vectors that are pairwise orthogonal. We also show that the limit of an optimal small d-codebook is an OMRS codebook. Both results demonstrate the necessity of OMRS for the optimality of small codebooks. We believe that OMRS is also a sufficient condition for optimality, but rather surprisingly, a formal proof seems difficult and will not be considered in this paper.

Finally, we observe that SRS is just a special case of OMRS for codebooks with cardinality equal to $R$. As a result, OMRS becomes the universal necessary condition for optimality.

Our results in this paper is in contrast to limited feedback beamforming in MISO systems, in which any set of linearly independent beamformers guarantee maximal diversity\cite{koyuncu2}, and the performance of a codebook is invariant under unitary transformations\cite{mukkavilli1, roh1, love1}. In that sense, this paper also shows that the vast literature on limited feedback beamforming for point-to-point systems is not directly applicable to cooperative networks, and we need new methods of analysis.

The rest of the paper is organized as follows: In Section \ref{secNetMod}, we introduce our system model, feedback and data transmission schemes, and problem definition. In Section \ref{secLowerBounds}, we present a fundamental lemma that we frequently use in our proofs. In Sections \ref{secSelecLarge} and \ref{secSelecSmall}, we state our main results on the necessity of SRS for large codebooks, and the necessity of OMRS for small codebooks, respectively. The numerical results are provided in Section \ref{secNumerical}. Finally, in Section \ref{secConc}, we draw our main conclusions. Some technical proofs are provided in the appendices.

\textit{Notation:} $||\cdot||_{\infty}$ indicates the infinite-norm. $\mathbb{C}$, $\mathbb{R}$, and $\mathbb{N}$ represent the sets of complex numbers, real numbers, and natural numbers, respectively. For $z\in\mathbb{C}$, $|z|$ indicates the absolute value. For a random variable $X$, $f_X(\cdot)$ and $F_X(\cdot)$ represent the probability density function (PDF) and the cumulative distribution function (CDF), respectively. $\mathcal{CN}(0, \sigma^2)$ represents a zero-mean complex Gaussian random variable with variance $\tfrac{\sigma^2}{2}$ per complex dimension. $\mathrm{E}[X]$ is the expected value of $X$. For any sets $\mathcal{A}$ and $\mathcal{B}$, $\mathcal{A}-\mathcal{B}$ is the set of elements in $\mathcal{A}$, but not in $\mathcal{B}$. $\mathcal{A}\subset \mathcal{B}$ means $\mathcal{A}$ is a subset of $\mathcal{B}$. $\mathcal{A}\cap\mathcal{B}$ and  $\mathcal{A}\cup\mathcal{B}$ are the intersection and the union of $\mathcal{A}$ and $\mathcal{B}$, respectively. $|\mathcal{A}|$ is the cardinality of $\mathcal{A}$, $\mathcal{A}^+ \triangleq \{x:x>0,\,x\in\mathcal{A}\}$, and $\mathcal{A}^r = \left\{(a_1,\ldots,a_r)|a_1,\ldots, a_r\in\mathcal{A}\right\}$, $r\in\mathbb{N}^+$ is the cartesian power. Finally, $\emptyset$ is the empty set, $\mathrm{Q}(\cdot)$ represents the Gaussian tail function, $\Gamma(\cdot)$ is the gamma function, $\log(\cdot)$ is the natural logarithm, and $\sinh(\cdot)$ is the hyperbolic sine.
\section{Network Model and Problem Statement}
\label{secNetMod}
\subsection{System Model}
\label{secSystemModel}
\begin{figure}[h]
\centering
\scalebox{1.75}{
\epsfig{file = 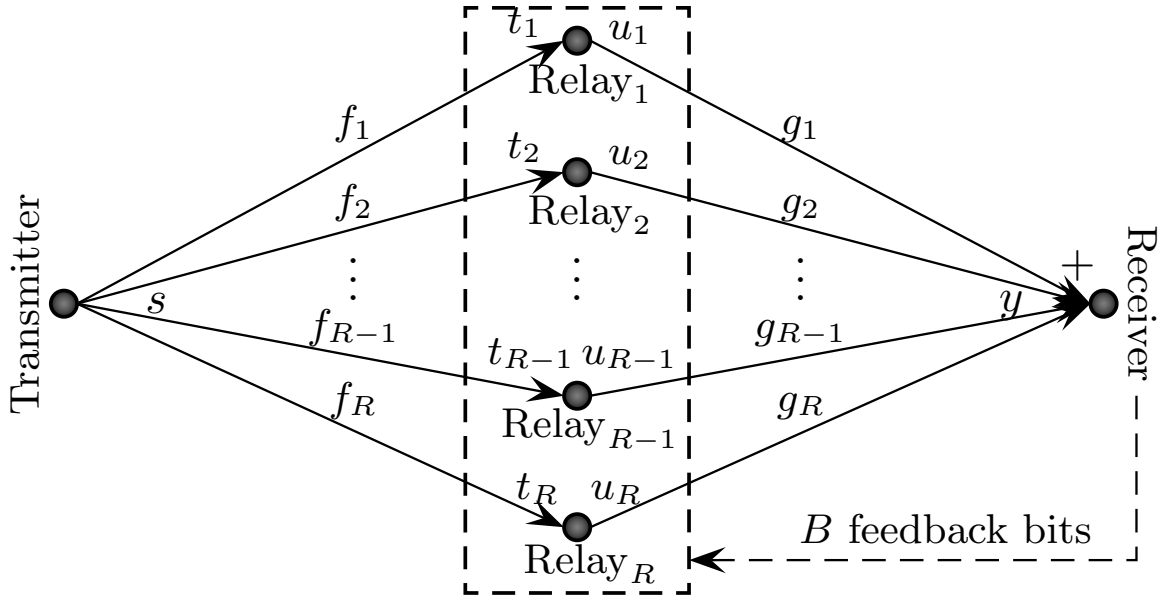, bbllx = 149pt, bblly = 597pt, bburx = 398pt, bbury = 725pt}}
\caption{System Block Diagram} \label{systemblockdiagram}
\end{figure}

The block diagram of the system is shown in Fig. \ref{systemblockdiagram}. We have a relay network with a transmitter-receiver pair and $R$ parallel relays. Let $f_r$ and $g_r$ denote the channel from the transmitter to the $r$th relay, and from the $r$th relay to the receiver, respectively. Also, let $\mathbf{h} = (f_1,g_1,\ldots,f_R,g_R)$ denote the channel state of the entire network. We assume that the entries of $\mathbf{h}$ are independent and $f_r \sim \mathcal{CN}(0, \sigma_{f_r}^2),\,g_r \sim \mathcal{CN}(0, \sigma_{g_r}^2),\,r=1,\ldots,R$.

Only the short-term power constraint is considered: For every symbol transmission, the average power levels used at the transmitter and the $r$th relay are no larger than $P_0$ and $P_r$, respectively. Let $P_i = p_i P,\,i=0,\ldots,R$, where $\infty>p_i>0$. In other words, we allow the power constraints of the transmitter and the relays to vary linearly with $P$. In addition, $P$ is the only network parameter that we allow to vary. All the remaining parameters (the channel variances $\sigma_{f_r},\sigma_{g_r},\,r=1,\ldots,R$ and the power constraint scalers $p_r,\,r=0,\ldots,R$) can be arbitrary, but we assume that they are fixed positive constants that do not depend on $P$.

We assume a quasi-static channel model, in which the channel realizations vary independently from one channel state to another, while within each state they remain constant. Also, we assume that the receiver knows the channel state of the entire network, $\mathbf{h}$; and the $r$th relay knows $|f_r|$. Each relay has also $B$ bits of partial CSI provided by receiver feedback.
\subsection{Feedback Transmission Scheme}
For any $P\in\mathbb{R}^+$ and a finite number of feedback bits $B$, the feedback transmission scheme operates as follows: For each frame, the channel realization $\mathbf{h}$ is quantized by a quantizer $\mathcal{Q}_P(\mathbf{h})\triangleq \mathtt{DEC}_P(\mathtt{ENC}_P(\mathbf{h}))$ defined by the encoder and decoder mappings $\mathtt{ENC}_P:\mathbb{C}^{2R}\rightarrow\mathcal{I}$, and $\mathtt{DEC}_P:\mathcal{I}\rightarrow\mathcal{D}_P$. In this definition, $\mathcal{D}_P$ represents the quantizer codebook for power level $P$, and $\mathcal{I}\triangleq \{1,\ldots,2^B\}$ denotes the index set for the codebook elements. We assume throughout the paper that codebooks for different power levels have the same cardinality, i.e. $|\mathcal{D}_P| = 2^B,\,\forall P$.

The encoding operation is performed at the receiver, and the feedback bits represent the encoder output. Each relay uses the decoder to find the corresponding codebook element. Each codebook element corresponds to a quantized beamforming vector. To summarize, we have a collection $\mathscr{Q} \triangleq \{\mathcal{Q}_P:P\in\mathbb{R}^+\}$ of quantizers. For a given $P\in\mathbb{R}^+$, we use the quantizer $\mathcal{Q}_P$ that provides the beamforming vector $\mathcal{Q}_P(\mathbf{h}) = \mathbf{x}$ for some $\mathbf{x}\in\mathcal{D}_P$.

Let $\mathcal{D}$ represent the mapping that maps a power level $P$ to its corresponding codebook $\mathcal{D}_P$. With some abuse of language, we call the set-valued map $\mathcal{D}$ a \textit{power-dependent codebook} (d-codebook) in the sense that for a given $P$, the codebook $\mathcal{D}_P$ is employed and can be optimized accordingly. Even though an optimal d-codebook can provide the best possible performance at any $P$, its accommodation requires the receiver and the relays to store a large number of codebooks. A more practical approach might be to use a \textit{power-independent codebook} (i-codebook) $\mathcal{C}$ as a special case of d-codebooks with $\mathcal{D}_P = \mathcal{C},\,\forall P$. We will have more to say on the practicality of i-codebooks later on.

%
%
\subsection{Data Transmission Scheme}
\label{secDataTransmisson}
We use a two-step AF protocol\cite{jing1, koyuncu1}. In the first step, the transmitter selects a symbol $s$ from a constellation $\mathcal{S}$, where $|\mathcal{S}|<\infty$, and sends $\sqrt{P_0}s$. We normalize $s$ as $\mathrm{E}[|s|^2]=1$. Thus, the average power used at the transmitter is $P_0$. During the first step, there is no reception at the receiver, but the $r$th relay receives
\begin{align}
t_r = f_r s\sqrt{P_0} + \eta_r,
\end{align}
where $\eta_{r}\sim\mathcal{CN}(0,1)$.

Suppose that $\mathcal{Q}_P(\mathbf{h}) = \mathbf{x}$, for some $\mathbf{x}\in\mathcal{D}_P$. Then, the relays use the beamforming vector $\mathbf{x}$ to adjust their transmit power and transmit phase. During the second step, the transmitters remain silent, but the $r$th relay transmits
\begin{align}
u_r = x_r \sqrt{\rho_r} t_r,
\end{align}
where
\begin{align}
\rho_r \triangleq \frac{P_r}{1 + |f_r|^2P_0}.
\end{align}

The average power used at the $r$th relay can be calculated to be $|x_r|^2P_r$. We require $0\leq|x_r|\leq 1$ as a result of the short term power constraint. The channel state dependent normalization factors $\rho_r$ ensure that the instantaneous transmit power of each relay remains within its power constraint with high probability.

Also, note that within the restriction of $0 \leq |x_r| \leq 1$, $\rho_r$ is the maximal normalization factor that we can use. In other words, if a factor $\rho_r''$ satisfies $\rho_r'' > \rho_r$ for some $\mathbf{h}$, then it violates the short term power constraint. Still, one can employ another factor $\rho_r'$ with $\rho_r' \leq \rho_r,\,\forall\mathbf{h}$ (e.g. $\rho_r' = P_r/(1+|f_r|^3P_0)$). We shall see later in Section \ref{secLowerBounds} that a different choice of the normalization factor does not change the main results of this paper.

After the two steps of transmission, the received signal at the receiver can be expressed as:
\begin{align}
   y =  \sum_{r=1}^R x_r\sqrt{\rho_r}f_{r}g_{r}\sqrt{P_0}s + \sum_{r=1}^R x_r g_{r}\sqrt{\rho_r} \eta_{r}  + \eta_{0},
\end{align}
where $\eta_{0}\sim\mathcal{CN}(0,1)$ is the noise at the receiver. We assume that the noises $\eta_i,\,i=0,\ldots,R$ and the channels are all independent. It follows that the received SNR is given by
\begin{align}
\label{receivedsnr}
 \mathtt{SNR}_P(\mathbf{x},\mathbf{h}) \triangleq \frac{P_0\left|\sum_{r=1}^R x_rf_rg_r\sqrt{\rho_r}\right|^2}{1 + \sum_{r=1}^R |x_r|^2|g_r|^2\rho_r}.
\end{align}
Since we have assumed that the power constraint scalers $p_i$ are fixed, the received SNR depends only on $P$ (indicated by a subscript), the beamforming vector $\mathbf{x}$, and the channel state $\mathbf{h}$.

In this work, our performance measure is the SER, and we present our results only for the case when $\mathcal{S} = \{+1,-1\}$ is a binary constellation. Then, the SER achieved by the quantizer $\mathcal{Q}_P$ at power level $P$ can be expressed as
\begin{align}
\label{serofaquantizer}
\mathtt{SER}_P(\mathcal{Q}_P) \triangleq E_{\mathbf{h}} [\mathrm{Q}\sqrt{2 \mathtt{SNR}_P(\mathcal{Q}_P(\mathbf{h}),\mathbf{h})}].
\end{align}
We would like to note that our results can be extended to any finite constellation $\mathcal{S}$.

Using (\ref{serofaquantizer}), the diversity achieved by the collection $\mathscr{Q}$ of quantizers is given by
\begin{align}
\label{divofaquantizer}
\diversity(\!\mathscr{Q}\;\!) \triangleq  \lim_{P\rightarrow\infty} -\frac{\log \mathtt{SER}_P(\mathcal{Q}_P)}{\log P}.
\end{align}
Since there are $R$ independently fading paths between the transmitter and the receiver, the maximal spatial diversity of our network model is $R$. In other words, for any $\mathscr{Q}$, $\diversity(\mathscr{Q}) \leq R$. A more formal proof of this argument can be found in \cite[Theorem 1]{koyuncu3}.
\subsection{Problem Statement}
\label{secProblemStatement}
Let $|\mathcal{D}|$ represent the common cardinality of each codebook $\mathcal{D}_P$. In other words, $|\mathcal{D}| = |\mathcal{D}_P| = 2^B,\,\forall P$.

We are interested in the structure of the optimal quantizers that minimize the SER subject to $|\mathcal{D}|= 2^B$. The following proposition from \cite{koyuncu1} determines the optimal quantizers given a fixed codebook.
\begin{proposition}
\label{optimalquantizerencoderlemma}
Given a fixed d-codebook $\mathcal{D}$ (i.e., for any $P$, the codebook $\mathcal{D}_P$ is fixed), the collection of optimal quantizers
that minimize the SER is given by $\mathscr{Q}_{\mathcal{D}}^{\star} \triangleq \{\mathcal{Q}_{P\!,\,\mathcal{D}_P}^{\star}:P\in\mathbb{R}^+ \}$, where
\begin{align}
\label{dcodebookoptimalencoder}
\mathcal{Q}_{P\!,\,\mathcal{D}_P}^{\star}(\mathbf{h}) \triangleq \arg\max_{\mathbf{x}\in\mathcal{D}_P} \mathtt{SNR}_P(\mathbf{x},\mathbf{h}),\,P\in\mathbb{R}^+.
\end{align}
In particular, given a fixed i-codebook $\mathcal{C}$, the collection of optimal quantizers is given by $\mathscr{Q}_{\mathcal{C}}^{\star} \triangleq \{\mathcal{Q}_{P\!,\,\mathcal{C}}^{\star}:P\in\mathbb{R}^+ \}$, where
\begin{align}
\label{icodebookoptimalencoder}
\mathcal{Q}_{P\!,\,\mathcal{C}}^{\star}(\mathbf{h}) = \arg\max_{\mathbf{x}\in\mathcal{C}} \mathtt{SNR}_P(\mathbf{x},\mathbf{h}),\,P\in\mathbb{R}^+.
\end{align}
\end{proposition}
In other words, for d-codebooks, given any power level $P$ and any fixed codebook $\mathcal{D}_P$ for $P$, the optimal quantizer encoder chooses the beamforming vector that maximizes the SNR at $P$. The interpretation of Proposition \ref{optimalquantizerencoderlemma} for an i-codebook $\mathcal{C}$ is analogous.

We would like to note that in practice, $|\mathcal{D}_P|<\infty$, and thus $\arg\max_{\mathbf{x}\in\mathcal{D}_P} \mathtt{SNR}_P(\mathbf{x},\mathbf{h})$ in (\ref{dcodebookoptimalencoder}) will always exist for any $\mathbf{h}$ and $P$. In order to be able to handle codebooks with $|\mathcal{D}_P| = \infty$, we shall further assume throughout the paper that $\mathcal{D}_P$ is compact for all $P$. Similarly, we assume that all i-codebooks are compact without explicit specification.

The main motivation for our introduction of i-codebooks was the claim that they are more practical than d-codebooks: One does not need to store different codebooks for different power levels. On the other hand, (\ref{icodebookoptimalencoder}) shows us that even if we use an i-codebook, the quantizer encoder will always depend on $P$. In that sense, one might argue that i-codebooks are as impractical as d-codebooks since a large number of quantizer encoders need to be stored anyway. Fortunately, for i-codebooks, we can observe from (\ref{icodebookoptimalencoder}) that the optimal encoder is a simple algebraic function of $P$. Therefore, we do not actually need to store the entire set of encoders. In order for a similar situation to hold for d-codebooks though, one needs a simple function that can map any power level to its corresponding optimal codebook. Finding such a function is an open problem. We thus present our results for both d-codebooks and i-codebooks, due to the potential optimality of the former and the practicality of the latter.
%

We shall use the optimal encoder in Proposition \ref{optimalquantizerencoderlemma} for the rest of the paper. Then, the codebook uniquely determines the performance of the system, and we set $\diversity(\mathcal{D}) \triangleq \diversity(\mathscr{Q}_{\mathcal{D}}^{\star})$ for a d-codebook $\mathcal{D}$, and similarly, $\diversity(\mathcal{C}) \triangleq \diversity(\mathscr{Q}_{\mathcal{C}}^{\star})$ for an i-codebook $\mathcal{C}$. Any optimal codebook should obey the following proposition from \cite{koyuncu1}.
\begin{proposition}
\label{codebookoptimalitylemma}
If $\mathcal{D}$ is an optimal d-codebook, then $\mathcal{D}_P\subset\mathcal{X},\,\forall P$, where $\mathcal{X} = \{\mathbf{x}:\mathbf{x}\in\mathbb{C}^R,\,\|\mathbf{x}\|_{\infty} = 1\}$. In particular, if $\mathcal{C}$ is an optimal i-codebook, then $\mathcal{C}\subset\mathcal{X}$.
\end{proposition}
In other words, at least one component of every beamforming vector in the codebook should have unit norm. Unless otherwise specified, we shall assume that all the codebooks in the rest of this paper are optimal in the sense of Proposition \ref{codebookoptimalitylemma}.

One simple, yet effective structured i-codebook is the SRS codebook, given by $\mathcal{C}_{\mathtt{SRS}}(\boldsymbol{\theta}) \triangleq \{\mathbf{e}_r(\theta_r):r=1,\ldots,R\}$, where $\boldsymbol{\theta} = \vect{\theta_1}{\cdots}{\theta_R}$, and $\mathbf{e}_r(\theta_r) \triangleq \vect{e_{r1}(\theta_r)}{\cdots}{e_{rR}(\theta_r)}$ with $e_{rq}(\theta_r)  = e^{j\theta_r},\,r=q$ and $\mathbf{e}_{rq}(\theta_r)  = 0,\,r\neq q$. As an example, both
\begin{align}
\label{zerolusrs}
\mathcal{C}_{\mathtt{SRS}}\left(\boldsymbol{0}\right) = \bigl\{\vect{0}{0}{1},\,\vect{0}{1}{0},\,\vect{1}{0}{0}\bigr\}
\end{align}
and
\begin{align}
\mathcal{C}_{\mathtt{SRS}}\left(\vect{\frac{\pi}{4}}{\frac{\pi}{2}}{\frac{2\pi}{3}}\right) = \bigl\{\vect{e^{j\frac{\pi}{4}}}{0}{0},\,\vect{0}{j}{0},\,\vect{0}{0}{e^{j\frac{2\pi}{3}}}\bigr\}
\end{align}
are SRS codebooks for a network with $3$ relays, where $\boldsymbol{0}$ represents the all-zero vector.

Even though there are infinitely many possible SRS codebooks given any $R$, all of them provide the same SER at any given $P$. This follows immediately from
\begin{proposition}
\label{snrinvariance}
For any beamforming vector $\mathbf{x}$ and channel state $\mathbf{h}$, we have
\begin{align}
\mathtt{SNR}_P(\mathbf{x}, \mathbf{h}) = \mathtt{SNR}_P(e^{j\theta}\mathbf{x}, \mathbf{h}),\,\forall\theta\in\mathbb{R}.
\end{align}
\end{proposition}
\begin{proof}
The proof is straightforward once we use the definition of $\mathtt{SNR}_P(\mathbf{x}, \mathbf{h})$ in (\ref{receivedsnr}).
\end{proof}

Let $\mathscr{C}_{\mathtt{SRS}} = \{\mathcal{C}_{\mathtt{SRS}}(\boldsymbol{\theta}):\boldsymbol{\theta}\in\mathbb{R}^R\}$ represent the collection of all possible SRS codebooks. It was shown in \cite[Theorem 1]{koyuncu1} that any $\mathcal{C}_{\mathtt{SRS}}\in\mathscr{C}_{\mathtt{SRS}}$ achieves the full-diversity order, $R$. Our first goal is to show that, in order to achieve diversity $R$, it is not only sufficient but also necessary to use SRS. To be more precise, it is necessary to include the SRS vectors to the quantizer codebook to achieve full-diversity. Since there are $R$ SRS vectors, any full-diversity codebook has thus cardinality at least $R$, and is necessarily large.

Clearly, we cannot choose the codebook cardinality freely at our will; given $B$ feedback bits, we are restricted to a codebook with cardinality $2^B$. As a result, one needs to use at least $\lceil \log_2 R\rceil$ bits of feedback to accommodate a large codebook and achieve full-diversity.

A low-rate application might require the number of feedback bits to be less than $\lceil \log_2 R\rceil$. In this case, we are restricted to using small codebooks and full-diversity is no longer achievable. Optimality conditions for small codebooks are more complicated than the ones for large codebooks and will be discussed later on.

All of our results on the necessity of relay selection will be based on a fundamental lemma that provides a lower bound on the SER of a given i-codebook. We introduce this lemma in the next section together with some example applications. We discuss the necessity of SRS for full-diversity immediately afterwards.
\section{Lower Bounds on the Performance of I-Codebooks}
\label{secLowerBounds}
We frequently use the following lemma to prove the main results in this paper. The proof of the lemma can be found in Appendix \ref{proofofmainlemma}.
\begin{lemma}
\label{mainlemma}
For any i-codebook $\mathcal{C}$, not necessarily with $\mathcal{C}\subset\mathcal{X}$, let 
\begin{align}
\label{indexsets}
\mathscr{R}(\mathcal{C}) \triangleq \{\mathcal{R}: \mathcal{R}\subset\{1,\ldots,R\}\mbox{ and }\forall \mathbf{x}\in\mathcal{C},\,\exists r\in\mathcal{R},\,x_r \neq 0\} 
\end{align}
be a collection of index sets. Then, there are constants $0 < \letpa,\,\letca,\,\letcb < \infty$ that are independent of $P$ and $\mathcal{C}$ s.t. for all $P \geq \letpa$ and $\mathcal{R}\in\mathscr{R}(\mathcal{C})$,
\begin{align}
\label{mainlemmaeq1}
\mathtt{SER}_P(\mathcal{Q}_{P\!,\,\mathcal{C}}^\star) \geq \letca [\xi(\mathcal{C},\mathcal{R})]^{\frac{3}{2}} \exp\left(-\frac{C_1}{\xi(\mathcal{C},\mathcal{R})}\right) P^{-|\mathcal{R}|},
\end{align}
where $\xi(\mathcal{C},\mathcal{R}) \triangleq \inf_{\mathbf{x}\in\mathcal{C}}\max_{r\in\mathcal{R}} |x_r|^2$.

Moreover, (\ref{mainlemmaeq1}) holds for any relay normalization factor $\rho_r' \leq \rho_r,\,\forall r$.
\end{lemma}

Since all of our main results will be based on the lower bound in (\ref{mainlemmaeq1}), and a different relay normalization factor will not improve this lower bound as stated in the lemma, we fix $\rho_r$ to be our relay normalization factor for the rest of the paper.

Before we discuss the consequences of Lemma \ref{mainlemma} regarding the necessity of relay selection, let us first present a motivating example application. As an immediate corollary to Lemma \ref{mainlemma}, the following theorem provides an upper bound on the diversity provided by any finite-cardinality i-codebook.
\begin{theorem}
\label{sidetheorem}
 For any i-codebook $\mathcal{C}$ with $|\mathcal{C}|<\infty$, $\diversity(\mathcal{C}) \leq \min\{|\mathcal{R}|:\mathcal{R}\in\mathscr{R}(\mathcal{C})\}$.
\end{theorem}
\begin{proof}
Since (\ref{mainlemmaeq1}) holds for any $\mathcal{R}\in\mathscr{R}(\mathcal{C})$, we choose the set $\mathcal{R}'$ in $\mathscr{R}(\mathcal{C})$ with the smallest cardinality (if the number of such sets is more than one, we can choose any of them). By definition, $\xi(\mathcal{C}, \mathcal{R}')$ is a positive constant that is independent of $P$. It follows from Lemma \ref{mainlemma} that
$\mathtt{SER}_P(\mathcal{Q}_{P,\mathcal{C}}^{\star}) \geq \letca [\xi(\mathcal{C},\mathcal{R}')]^{3/2} \exp(-C_1/\xi(\mathcal{C},\mathcal{R}')) P^{-|\mathcal{R}'|},\,\forall P \geq \letpa$. Thus, $\mathcal{C}$ provides at most a diversity of $|\mathcal{R}'|$.
\end{proof}

The rest of this section is devoted to some example applications of this theorem.

\begin{example}\label{example1}\emph{
For a network with $3$ relays, let
\begin{align}
\label{cbc1} \mathcal{C}_1 & = \bigl\{\vect{0}{1}{1}\bigr\}, \\
\label{cbc2} \mathcal{C}_2 & = \bigl\{\vect{0}{1}{1},\,\vect{1}{0}{1}\bigr\},\\
\label{cbc3} \mathcal{C}_3 & = \bigl\{\vect{0}{1}{1},\,\vect{1}{0}{1},\,\vect{1}{1}{0}\bigr\}.
\end{align}
Let us first find an upper bound on the diversity provided by $\mathcal{C}_1$. Using the definition in (\ref{indexsets}), we have $\mathscr{R}(\mathcal{C}_1) = \{\{2\},\{3\},\{1,2\},\{2,3\},\{1,3\},\{1,2,3\}\}$.
Then, according to Theorem \ref{sidetheorem}, $\diversity(\mathcal{C}_1) \leq 1$ since
\begin{align}
\min\{|\mathcal{R}|:\mathcal{R}\in\mathscr{R}(\mathcal{C}_1)\} & = \min\{|\{2\}|,|\{3\}|,|\{1,2\}|,|\{2,3\}|,|\{1,3\}|,|\{1,2,3\}|\} \\ & = \min\{1,1,2,2,2,3\} \\ & = 1.
\end{align}
Similarly, $\diversity(\mathcal{C}_2) \leq 1$ since $\mathscr{R}(\mathcal{C}_2) = \{\{3\},\{1,2\},\{2,3\},\{1,3\},\{1,2,3\}\}$ and thus $\min\{|\mathcal{R}|:\mathcal{R}\in\mathscr{R}(\mathcal{C}_2)\} = 1$. On the other hand, the ``best'' that we can say about the diversity of $\mathcal{C}_3$ is that $\diversity(\mathcal{C}_3) \leq 2$ since $\mathscr{R}(\mathcal{C}_3) = \{\{1,2\},\{2,3\},\{1,3\},\{1,2,3\}\}$ with $\min\{|\mathcal{R}|:\mathcal{R}\in\mathscr{R}(\mathcal{C}_3)\} = 2$.\qed}
\end{example}
\begin{example}\label{example2}\emph{
None of the codebooks $\mathcal{C}_1, \mathcal{C}_2$ and $\mathcal{C}_3$ in Example \ref{example1} can achieve the maximal diversity order $3$. Now, suppose that a finite-cardinality i-codebook $\mathcal{C}_4$ achieves diversity $3$. Then, we should have $\{1,2\},\{1,3\},\{2,3\}\notin\mathscr{R}(\mathcal{C}_4)$ (otherwise, if e.g., $\{1,2\}\in\mathscr{R}(\mathcal{C}_4)$, then according to Theorem \ref{sidetheorem}, $\diversity(\mathcal{C}_4) \leq 2$). Now, since $\{1,2\}\notin\mathscr{R}(\mathcal{C}_4)$, by the definition of $\mathscr{R}(\cdot)$ in (\ref{indexsets}), $\exists\mathbf{x}=[x_1\, x_2 \,x_3]\in\mathcal{C}_4$ s.t. $|x_1| = |x_2| = 0$. Also, as a result of Proposition \ref{codebookoptimalitylemma}, $|x_3| = 1$, and thus $\exists\theta_3\in\mathbb{R}$ s.t. $x_3 = e^{j\theta_3}$.  In other words, $\mathbf{x} = \mathbf{e}_3(\theta_3)$ is an SRS vector. Similarly, using the conditions that $\{1,3\}\notin\mathscr{R}(\mathcal{C}_4)$ and $\{2,3\}\notin\mathscr{R}(\mathcal{C}_4)$, we can show that $\exists\theta_2\in\mathbb{R},\,\mathbf{e}_2(\theta_2)\in\mathcal{C}_4$ and $\exists\theta_1\in\mathbb{R},\,\mathbf{e}_1(\theta_1)\in\mathcal{C}_4$, respectively. Therefore, only if $\exists\mathcal{C}_{\mathtt{SRS}}\in\mathscr{C}_{\mathtt{SRS}}$ s.t. $\mathcal{C}_{\mathtt{SRS}}\subset\mathcal{C}_4$, we can have $\diversity(\mathcal{C}_4) = 3$. But, we also know from \cite{koyuncu1} that $\forall\mathcal{C}_{\mathtt{SRS}}\in\mathscr{C}_{\mathtt{SRS}}$, if $\mathcal{C}_{\mathtt{SRS}}\subset\mathcal{C}_4$, we have $\diversity(\mathcal{C}_4) = 3$. Hence, for a network with $3$ relays, a finite-cardinality i-codebook can achieve diversity $3$ \textit{if and only if} it contains an SRS codebook. In Section \ref{secSelecLarge}, we shall generalize this result to networks with any number of relays that employ codebooks with possibly infinite cardinality.}\qed
\end{example}
\begin{example}[Comparison with MISO Systems]\label{example3}\emph{
One of the most surprising conclusions that we can draw from Theorem \ref{sidetheorem} is that, unlike a MISO system, in a relay network, (i) the performance of a codebook can significantly vary under unitary transformations, and (ii) the existence of linearly independent codebook vectors do not guarantee maximal diversity. We have demonstrated the latter phenomenon by codebooks $\mathcal{C}_2$ and $\mathcal{C}_3$ in Example \ref{example1}. Despite the fact that $\mathcal{C}_2$ and $\mathcal{C}_3$ consist of $2$ and $3$ linearly independent codebook vectors, respectively, we have $\diversity(\mathcal{C}_2) \leq 1$ and $\diversity(\mathcal{C}_3) \leq 2$.}

\emph{We now demonstrate the former phenomenon. For that purpose, let $\mathcal{C}\cdot\mathbf{U} \triangleq \{\mathbf{x}\mathbf{U} : \mathbf{x}\in\mathcal{C}\}$ denote the transformation of the codebook $\mathcal{C}$ by a unitary matrix $\mathbf{U}$.}

\emph{In this example, we consider networks with a sum-power constraint $P$ on relays. For such networks, the $r$th relay transmits with power $|x_r|^2P$ given a beamforming vector $\mathbf{x}$, and we require $\sum_{r=1}^R |x_r|^2 P \leq P \implies \|\mathbf{x}\| \leq 1$. The sum-power constraint on relays makes sure that if $\mathcal{C}$ is a feasible codebook, then for any unitary matrix $\mathbf{U}$, the codebook $\mathcal{C}\cdot\mathbf{U}$ is also feasible.}

\emph{Let us now consider the transformations of the codebook $\mathcal{C}_{\mathtt{SRS}}(\boldsymbol{0})$ in (\ref{zerolusrs}) by the unitary matrices
\begin{align}
\label{someunitaries}
\mathbf{U}_1 = \left[\begin{array}{ccc} \frac{1}{\sqrt{2}} & \frac{1}{\sqrt{2}} & 0 \\ -\frac{j}{\sqrt{2}} & \frac{j}{\sqrt{2}} & 0 \\ 0 & 0 & j \end{array}\right],\mbox{ and }
 \mathbf{U}_2 = \frac{1}{4}\left[\begin{array}{ccc} 1+j & 1-3j & -\sqrt{2}+j\sqrt{2} \\ -3-j & 1-j & \sqrt{2}+j\sqrt{2} \\ \sqrt{2}+j\sqrt{2} & \sqrt{2}+j\sqrt{2} & 2 + 2j \end{array}\right].
\end{align}
Note that the codebooks $\mathcal{C}_{\mathtt{SRS}}(\boldsymbol{0}) \cdot \mathbf{U}_1$ and $\mathcal{C}_{\mathtt{SRS}}(\boldsymbol{0}) \cdot \mathbf{U}_2$ consist of the rows of $\mathbf{U}_1$ and $\mathbf{U}_2$, respectively.}

\emph{For limited feedback MISO systems with independent and identically distributed transmitter-to-receiver channels\cite{roh1,love1,mukkavilli1}, the performance of a quantizer codebook is invariant under unitary transformations. Moreover, even in the case of arbitrary channel variances, the diversity of a codebook is preserved under unitary transformations.
On the other hand, for our example network, the application of Theorem \ref{sidetheorem} yields $\mathtt{d}(\mathcal{C}_{\mathtt{SRS}}(\boldsymbol{0}) \cdot \mathbf{U}_1) \leq 2$, and $\mathtt{d}(\mathcal{C}_{\mathtt{SRS}}(\boldsymbol{0}) \cdot \mathbf{U}_2) \leq 1$, whereas $\diversity(\mathcal{C}_{\mathtt{SRS}}(\boldsymbol{0})) = 3$. In general, unless $\mathbf{U}$ is diagonal, it can be shown that $\diversity(\mathcal{C}_{\mathtt{SRS}}(\boldsymbol{0}) \cdot \mathbf{U}) \leq 2$. Therefore, in relay networks, even the diversity performance of a codebook is not preserved under unitary transformations. This unexpected behavior can be attributed to the non-linear nature of the distortion function as well as the noise amplification at the relays.}\qed
\end{example}

\section{The Necessity of SRS}
\label{secSelecLarge}
With Lemma \ref{mainlemma} at hand, we can now introduce our results on the necessity of relay selection. In this section in particular, we determine the structure of optimal quantizers that achieve the full-diversity order, $R$. First, we consider the power-independent i-codebooks, and show that every i-codebook that achieves full-diversity necessarily contains the SRS codebook. We then focus on d-codebooks that can be optimized with respect to the power level $P$, and show that an optimal large d-codebook contains an SRS codebook asymptotically as $P$ grows to infinity.
\subsection{The Necessity of SRS - I-Codebooks}
In Example \ref{example2} in Section \ref{secLowerBounds}, we showed that an i-codebook for a network with $3$ relays can achieve full-diversity if and only if it contains an SRS codebook. The following theorem generalizes this result to networks with any number of relays that employ codebooks with possibly infinite cardinality.
\begin{theorem}
\label{litheorem}
For any i-codebook $\mathcal{C}$, $\diversity(\mathcal{C})= R$ if and only if $\exists\mathcal{C}_{\mathtt{SRS}}\in\mathscr{C}_{\mathtt{SRS}}$ s.t. $\mathcal{C}_{\mathtt{SRS}} \subset \mathcal{C}$.
\end{theorem}
\begin{proof}
The ``if'' part was proved in \cite{koyuncu1}. Here, we prove the ``only if'' part by contradiction. Suppose there is a compact i-codebook $\mathcal{C}$ with  $\diversity(\mathcal{C}) = R$ and $\forall\mathcal{C}_{\mathtt{SRS}}\in\mathscr{C}_{\mathtt{SRS}}$, $\mathcal{C}_{\mathtt{SRS}}$ is not a subset of $\mathcal{C}$. The latter condition implies that $\exists r\in\{1,\ldots,R\},\,\forall\theta_r,\,\mathbf{e}_r(\theta_r) \notin \mathcal{C}$ (as otherwise, $\forall r\in\{1,\ldots,R\},\,\exists\vartheta_r,\,\mathbf{e}_r(\vartheta_r) \in \mathcal{C}$ and thus $\mathcal{C}_{\mathtt{SRS}}([\vartheta_1\cdots\vartheta_R])\subset\mathcal{C}$). In other words, $\mathcal{C}$ does not contain the vector(s) that selects the $r$th relay.

Let $\mtb \triangleq \{\mathbf{e}_r(\theta):\theta\in\mathbb{R}\}$. Note that $\mtb$ is the product of the closure of the unit disk by the all-zero vector of dimension $R-1$. Since all the factor sets are compact, $\mtb$ is compact.

We now show that
\begin{align}
\label{firstcondition}
\exists \epsilon > 0,\,\forall\mathbf{t}\in\mtb,\,\forall\mathbf{x}\in\mathcal{C},\,\|\mathbf{x} - \mathbf{t}\| > \epsilon.
\end{align}
Let $(\mathbf{x}',\mathbf{t}') = \arg\min_{(\mathbf{x},\mathbf{t})\in\mathcal{C}\times\mtb} \|\mathbf{x} - \mathbf{t}\|$. The minimum will always exist as $\mathcal{C}\times\mtb$ is compact and $f(\mathbf{x},\mathbf{t}) = \|\mathbf{x}-\mathbf{t}\|$ is continuous. Moreover, since $\mathcal{C}\cap\mtb=\emptyset$, we have $\mathbf{x}'\notin\mtb$, and thus $\|\mathbf{x}'-\mathbf{t}'\| > 0$. Therefore, we can pick e.g. $\epsilon = \frac{1}{2} \|\mathbf{x}'-\mathbf{t}'\| > 0$, and (\ref{firstcondition}) will hold.

According to (\ref{firstcondition}), for any $\mathbf{x}\in\mathcal{C}$, we have
\begin{align}
\label{ycond}
 |x_r - e^{j\theta}|^2 + \sum_{\substack{q=1 \\ q\neq r}}^R |x_q|^2 > \epsilon^2,\,\forall \theta\in\mathbb{R}. \end{align}
Also, since $\mathbf{x}\in\mathcal{X}$, it follows that $\exists r'\in\mathcal{R},\,|x_{r'}| = 1$. If $r' = r$, we choose $\theta' = \angle x_{r'}$. Then, (\ref{ycond})$\implies\sum_{q\neq r} |x_q|^2 > \epsilon^2 \implies \max_{q\neq r} |x_q|^2 > (R-1)^{-1}\epsilon^2$. Otherwise, if $r'\neq r$, then $\exists q\neq r,\,|x_q| = 1 \implies \max_{q\neq r} |x_q|^2 = 1$. In either case, $\max_{q\neq r} |x_q|^2 > \epsilon_0$, where $\epsilon_0\triangleq \min\{1, (R-1)^{-1}\epsilon^2\} > 0$.

Now, let $\mathcal{U} = \{1,\ldots,R\} - \{r\}$. Clearly, $\mathcal{U}\in\mathscr{R}(\mathcal{C})$. Moreover, $\xi(\mathcal{C},\mathcal{U}) = \inf_{\mathbf{x}\in\mathcal{C}} \max_{q\neq r} |x_q|^2 \geq \epsilon_0$. Using Lemma \ref{mainlemma}, $\mathtt{SER}_P(\mathcal{Q}^{\star}_{P,\mathcal{C}}) \geq \letca \epsilon_0^{3/2}\exp(-C_1/\epsilon_0)P^{-|\mathcal{U}|},\,\forall P \geq \letpa$. Therefore, $\diversity(\mathcal{C}) \leq |\mathcal{U}| = R-1$, which contradicts the assumption that $\diversity(\mathcal{C}) = R$. This concludes the proof.
\end{proof}
Therefore, an i-codebook can achieve diversity $R$ if and only if it contains an SRS codebook.
\subsection{The Necessity of SRS - D-Codebooks}
Let us now consider the necessity of SRS for power-dependent d-codebooks. In this paper, we are interested in the asymptotic structure of optimal d-codebooks as $P$ grows to infinity. As we have mentioned in Section \ref{secDataTransmisson}, we can interpret any d-codebook $\mathcal{D}$ as a set-valued map that maps the power level $P\in\mathbb{R}^+$ to the codebook $\mathcal{D}_P\subset\mathcal{X}\subset\mathbb{C}^R$. Therefore, we use the well-established limit definitions for set-valued maps\cite{sva} to characterize the asymptotic structure (or the limit) of any d-codebook.
\begin{definition}[See, e.g. {\cite[Definition 1.4.6]{sva}}]
Let $\mathcal{D}$ be a d-codebook. For any $P$ and $\mathbf{x}\in\mathbb{C}^R$, let
\begin{align}
\label{compactminimumlanbnu}
d_P(\mathbf{x}) & \triangleq \min_{\mathbf{y}\in\mathcal{D}_P} \|\mathbf{x} - \mathbf{y}\|
\end{align}
as the distance of $\mathbf{x}$ to $\mathcal{D}_P$. The minimum in (\ref{compactminimumlanbnu}) always exists since $\mathcal{D}_P$ is compact for all $P$.

We now define
\begin{align}
\limsup_{P \rightarrow\infty} \mathcal{D}_P \triangleq \left\{\mathbf{x}\in\mathbb{C}^R:\liminf_{P\rightarrow\infty} d_P(\mathbf{x}) = 0\right\}
\end{align}
as the \textit{upper limit} of $\mathcal{D}_P$ as $P \rightarrow \infty$, and
\begin{align}
\liminf_{P \rightarrow\infty} \mathcal{D}_P \triangleq \left\{\mathbf{x}\in\mathbb{C}^R:\lim_{P\rightarrow\infty} d_P(\mathbf{x}) = 0\right\}
\end{align}
as the \textit{lower limit} of $\mathcal{D}_P$ as $P \rightarrow \infty$. The upper and lower limits always exist for any given $\mathcal{D}$.

If $\liminf_{P \rightarrow\infty} \mathcal{D}_P = \limsup_{P \rightarrow\infty} \mathcal{D}_P = \mathcal{L}$, i.e. if the upper and lower limits agree, we say that the d-codebook converges to $\mathcal{L}$ and write $\lim_{P\rightarrow\infty}\mathcal{D}_P = \mathcal{L}$.

We also use the shorthand notation $\liminf \mathcal{D} \triangleq \liminf_{P\rightarrow\infty}\mathcal{D}_P$, $\limsup\mathcal{D} \triangleq \limsup_{P\rightarrow\infty}\mathcal{D}_P$, and similarly, $\lim\mathcal{D}  \triangleq \lim_{P\rightarrow\infty}\mathcal{D}_P $.
\end{definition}

Given Theorem \ref{litheorem}, we expect intuitively that the limit of any full-diversity d-codebook $\mathcal{D}$ contains an SRS codebook provided that $\lim\mathcal{D}$ exists. The following theorem, whose proof can be found in Appendix \ref{proofofldtheorem}, verifies this intuition:
\begin{theorem}
\label{ldtheorem}
The following arguments hold for any d-codebook $\mathcal{D}$ with $\diversity({\mathcal{D}}) = R$.
\begin{enumerate}
\item There are $R$ distinct beamforming vectors $\widetilde{\mathbf{e}}_{r,P},\,r=1,\ldots,R$ in $\mathcal{D}_P$ s.t. for all $P>\letpththree$,
\begin{align}
\label{rateofconv}
|\widetilde{e}_{r,P,q}|^2 \leq \frac{\ththreeconst}{\log P},\,\forall q\in\{1,\ldots,R\}-\{r\},\,r=1,\ldots,R,
\end{align}
where $\widetilde{e}_{r,P,q}$ represents the $q$th component of $\widetilde{\mathbf{e}}_{r,P}$, and $0<\letpththree, \ththreeconst<\infty$ are constants that are independent of $P$ and $\mathcal{D}$.
\item If $\lim\mathcal{D}$ exists,
\begin{enumerate}
\item If $|\mathcal{D}| = R$, $\exists\mathcal{C}_{\mathtt{SRS}}\in\mathscr{C}_{\mathtt{SRS}}$ s.t.  $\mathcal{C}_{\mathtt{SRS}}=\lim \mathcal{D}$.
\item If $|\mathcal{D}| > R$, $\exists\mathcal{C}_{\mathtt{SRS}}\in\mathscr{C}_{\mathtt{SRS}}$ s.t.  $\mathcal{C}_{\mathtt{SRS}}\subset \lim \mathcal{D}$.
\end{enumerate}
\item If $\lim \mathcal{D}$ does not exist, $\exists\mathcal{C}_{\mathtt{SRS}}\in\mathscr{C}_{\mathtt{SRS}}$ s.t.  $\mathcal{C}_{\mathtt{SRS}}\subset \limsup \mathcal{D}$.
\end{enumerate}
\end{theorem}

Since we can achieve full-diversity using the SRS scheme, any SER-optimal large d-codebook should achieve full-diversity as well. In that sense, the necessary conditions that we have stated in Theorem \ref{ldtheorem} hold for optimal large d-codebooks as well.

Also note that the rate of convergence indicated in (\ref{rateofconv}) is only a necessary condition. In other words, a sequence of codebooks satisfying (\ref{rateofconv}) do not necessarily provide maximal diversity. We conjecture that a necessary and sufficient rate of convergence is $\frac{1}{P}$ instead of $\frac{1}{\log P}$ stated in the theorem.

In Theorem \ref{ldtheorem}, we have also taken into account codebooks that may fail to converge. This is not a limitation of the analysis that has been carried out: There exists optimal d-codebooks that do not converge as we demonstrate by the following proposition:
\begin{proposition}
For any $R$ and $B <\infty$, there exists an optimal d-codebook $\mathcal{D}$ with $|\mathcal{D}| = 2^B$ that does not converge.
\end{proposition}
\begin{proof}
We prove the proposition for the trivial case $R=1$ and give a sketch of the proof for $R > 1$. For $R=1$, the received SNR is given by 
\begin{align}
\mathtt{SNR}_P(\mathbf{x},\mathbf{h}) = \frac{|x_1|^2|f_1|^2|g_1|^2 P_0P_1}{1 + |f_1|^2P_0 + |g_1|^2P_1}. 
\end{align}
Hence, at any $P$, it is sufficient to use a single beamforming ``vector'' $[x]$ with $|x| = 1$ to achieve the best SER performance; it is needless to use a codebook with cardinality greater than $1$. As an example, a d-codebook $\mathcal{D}$ with $\mathcal{D}_P = \{[1]\},\,\forall P$ is SER-optimal.

Let us now define another d-codebook $\mathcal{D}'$ as $\mathcal{D}_P' = \{[-1]\}$ if $n \leq P < n+1$ for some nonnegative integer $n$, and $\mathcal{D}_P' = \{[1]\}$, otherwise. Note that $\mathcal{D}'$ provides the same SER performance as $\mathcal{D}$. On the other hand, it is straightforward to show that $\limsup \mathcal{D}' = \{[1],\,[-1]\}$, and $\liminf \mathcal{D}' = \emptyset$, and hence $\lim\mathcal{D}'$ does not exist.

In general, for any $R$ and a finite $B$, we can synthesize a non-convergent optimal d-codebook out of a convergent optimal d-codebook $\mathcal{D}$ as follows: If $n \leq P < n+1$ for some nonnegative integer $n$, we replace a beamforming vector $\mathbf{x}\in\mathcal{D}_P$ by $e^{j\theta}\mathbf{x}$ for some $\theta\in\mathbb{R}$, and otherwise, leave it unchanged. As a result of Proposition \ref{snrinvariance}, the new d-codebook provides the same performance as $\mathcal{D}$, and is thus optimal. However, it fails to have a limit due to the artificial phase oscillations that we have introduced.
\end{proof}
In this section, we have shown by Theorems \ref{litheorem} and \ref{ldtheorem} that one needs to include all the SRS vectors to the quantizer codebook to achieve full-diversity. This requires the accommodation of a large codebook, or equivalently, at least $\lceil \log_2 R \rceil$ bits of feedback. On the other hand, the design constraints might require that the number of available feedback bits is less than $\lceil \log_2 R \rceil$, in which case we are restricted to using small codebooks and full-diversity is no longer achievable. Our goal in the next section is to determine the optimal codebook structure for such low feedback rate applications.
\section{Small Codebooks and the Necessity of OMRS}
\label{secSelecSmall}
In this section, we first determine the maximal achievable diversity with small codebooks. Then, we find the optimal small codebook structure that can achieve maximal diversity. We show that a diversity-optimal small i-codebook should contain multiple-relay selection vectors that are pairwise orthogonal, i.e. it should be an OMRS codebook. We also demonstrate the necessity of OMRS for small d-codebooks. Finally, we observe that SRS is actually a special case of OMRS for codebooks with cardinality equal to $R$. Therefore, OMRS becomes the universal necessary condition for codebook optimality.
\subsection{Diversity Limitations of Small Codebooks}
The following theorem shows that the maximal diversity provided by any small codebook is equal to the cardinality of the codebook.
\begin{theorem}
\label{smallmaxdiversitytheorem}
For any d-codebook $\mathcal{D}$, $\diversity(\mathcal{D}) \leq \min\{R, |\mathcal{D}|\}$.
\end{theorem}
\begin{proof}
Let us first prove the theorem for an i-codebook $\mathcal{C}$. For any $\mathbf{x}\in\mathcal{X}$, let $\iota(\mathbf{x})\in\{1,\ldots,R\}$ be any index with $|x_{\iota(\mathbf{x})}| = 1$. Note that as a result of Proposition \ref{codebookoptimalitylemma}, $\iota(\mathbf{x})$ always exists whenever $\mathbf{x}$ is a member of a quantizer codebook.

Now let $\mtc = \{\iota(\mathbf{x}):\mathbf{x}\in\mathcal{C}\}$. Note that $|\mtc| \leq \min\{R, |\mathcal{C}|\}$, $\mtc\in\mathscr{R}(\mathcal{C})$, and $\xi(\mathcal{C}, \mtc) =1$. Using Lemma \ref{mainlemma}, $\forall P \geq \letpa$, we have
\begin{align}
\mathtt{SER}_P(\mathcal{Q}_{P, \mathcal{C}}^{\star}) & \geq \thfourconst P^{-|\mtc|} \\
\label{smalllowboundd}& \geq \thfourconst P^{-\min\{R, |\mathcal{C}|\}},
\end{align}
where $\thfourconst \triangleq \letca\exp(-C_1)$. Thus, $\diversity(\mathcal{C})\leq \min\{R, |\mathcal{C}|\}$, concluding the proof for i-codebooks.

One way to deal with the complications that arise from the power-dependency of d-codebooks is to define a lower bound that treats each codebook $\mathcal{D}_{\varrho},\,\varrho\in\mathbb{R}^+$ as an i-codebook. At a given $P$, we can calculate the SERs of all $\mathcal{D}_{\varrho},\,\varrho\in\mathbb{R}^+$. The infimum of these SERs then gives us a lower bound on the performance of $\mathcal{D}$ at $P$. With this observation, $\forall P \geq \letpa$, we have
\begin{align}
\mathtt{SER}_P(\mathcal{Q}_{P\!,\,\mathcal{D}_P}^\star) & \geq \inf_{\mathcal{D}_{\varrho}:\varrho\in\mathbb{R}^+} \mathtt{SER}_P(\mathcal{Q}_{P\!,\,\mathcal{D}_{\varrho}}^\star) \\
\label{infimumfun} & \geq \inf_{\mathcal{D}_{\varrho}:\varrho\in\mathbb{R}^+} \thfourconst P^{-\min\{R, |\mathcal{D}|\}} \\
& = \thfourconst P^{-\min\{R, |\mathcal{D}|\}},
\end{align}
where (\ref{infimumfun}) follows from (\ref{smalllowboundd}). This concludes the proof.
\end{proof}
There are structured small codebooks that can achieve the diversity upper bound in Theorem \ref{smallmaxdiversitytheorem}. As an example, for an i-codebook $\mathcal{C}_{\mathtt{SRS}}'(d,\boldsymbol{\theta}) = \{\mathbf{e}_i(\theta_i),\,i=1,\ldots,d\}$ that contains $d < R$ SRS vectors, $\diversity(\mathcal{C}_{\mathtt{SRS}}'(d,\boldsymbol{\theta})) = d,\,\forall\boldsymbol{\theta}$, as shown in \cite{koyuncu1}. In other words, an ``incomplete'' SRS scheme, in which the selection of only a subset of the relays is considered, can achieve maximal diversity. What is left is thus to determine the structure of a general diversity-optimal small codebook. Unlike large codebooks where SRS is the only way to achieve maximal diversity, we show in the following that for small codebooks, a more general OMRS structure can potentially provide maximal diversity.
\subsection{OMRS}
The necessity of SRS for large codebooks ``generalizes'' to the necessity of OMRS for small codebooks. Let us first describe what we mean by OMRS in a more formal manner.
\begin{definition}[OMRS] An i-codebook $\mathcal{C}$ is an OMRS codebook if and only if either $|\mathcal{C}| = 1$, or $\forall\mathbf{x},\mathbf{y}\in\mathcal{C},\,\mathbf{y}\neq\mathbf{x},\,\sum_{r=1}^R |x_r||y_r| = 0$. OMRS is the scheme induced by an OMRS codebook.
\end{definition}

In other words, an OMRS codebook contains multiple-relay selection vectors that are pairwise orthogonal.\footnote{Note that this orthogonality condition is not the same as the ``usual'' orthogonality condition for complex vectors with respect to the Hermitian inner product.} As an example, $\mathcal{C}_5 = \{[0\;1\;0\;0.8], [0\;0\;1\;0], [1\;0\;0\;0]\}$ is an OMRS codebook.

By definition, the cardinality of an OMRS codebook cannot be more than $R$. An OMRS codebook that has cardinality equal to $R$ should be familiar: it is an SRS codebook.
\subsection{The Necessity of OMRS - I-Codebooks}
Now let us demonstrate the necessity of OMRS for i-codebooks by the following theorem:
\begin{theorem}
\label{sitheorem}
A diversity-optimal i-codebook $\mathcal{C}$ with $|\mathcal{C}| \leq R$ is an OMRS codebook.
\end{theorem}
\begin{proof}
The case $|\mathcal{C}|=1$ is trivial. We prove the other cases by contradiction.
Suppose that there exists a non-OMRS i-codebook $\mathcal{C}$ with $1 < |\mathcal{C}| \leq R$ and $\diversity(\mathcal{C}) = |\mathcal{C}|$. Since $\mathcal{C}$ is not an OMRS, $\exists\mathbf{x},\mathbf{y}\in\mathcal{C},\,\mathbf{y}\neq\mathbf{x},\,\exists r\in\{1,\ldots,R\},\,|x_r| \neq 0 ,\,|y_r|\neq 0$. Now, let $\mtd = \{r\}\cup\{\iota(\mathbf{z}): \mathbf{z}\in\mathcal{C} - \{\mathbf{x},\mathbf{y}\}\}$, where $\iota(\mathbf{z})$ is any index that satisfies $|z_{\iota(\mathbf{z})}| = 1$. Note that $|\mtd| \leq |\mathcal{C}| - 1$, $\mtd\in\mathscr{R}(\mathcal{C})$, and $\xi(\mathcal{C}, \mtd) = \min\{|x_r|, |y_r|\}$. Applying Lemma \ref{mainlemma}, we have $\diversity(\mathcal{C}) \leq |\mathcal{C}| - 1$. This contradicts the assumption that $\diversity(\mathcal{C}) = |\mathcal{C}|$.
\end{proof}
In other words, an i-codebook $\mathcal{C}$ with $|\mathcal{C}| \leq R$ achieves diversity $|\mathcal{C}|$ only if it is an OMRS codebook. In particular, if $|\mathcal{C}| = R$, $\diversity(\mathcal{C}) = R$ if and only if $\mathcal{C}$ is an OMRS codebook, in which case it is also an SRS codebook. Unlike the necessity and sufficiency of SRS for large codebooks, we can only show the necessity of OMRS for small codebooks. We leave the sufficiency as a conjecture:
\begin{conjecture}
If a small i-codebook $\mathcal{C}$ is an OMRS codebook, $\diversity(\mathcal{C}) = |\mathcal{C}|$.
\end{conjecture}
\subsection{The Necessity of OMRS - D-Codebooks}
Let us now generalize our result on the necessity of OMRS for i-codebooks to d-codebooks by the following theorem. Its proof can be found in Appendix \ref{proofofsdtheorem}.
\begin{theorem}
\label{sdtheorem}
Let $\mathscr{O}(c)$ denote the collection of all possible OMRS codebooks with cardinality $c$. The following arguments hold for any optimal d-codebook $\mathcal{D}$ with $1 \leq |\mathcal{D}| \leq R$.
\begin{enumerate}
\item There are constants $0<\sdtheoremcb,\,\sdtheorempb < \infty$ that are independent of $P$ and $\mathcal{D}$ s.t. for all $P > \sdtheorempb$,
\begin{align}
\label{sdtheoremeq}
\max_{\substack{\mathbf{x},\mathbf{y}\in\mathcal{D}_P \\ \mathbf{x}\neq\mathbf{y}}} \sum_{r=1}^R |x_r||y_r| \leq \frac{\sdtheoremcb}{\log P}.
\end{align}
\item If $\lim \mathcal{D}$ exists, $\exists\mathcal{O}\in\mathscr{O}(|\mathcal{D}|)$ s.t. $\mathcal{O}= \lim \mathcal{D}$.
\item If $\lim \mathcal{D}$ does not exist, $\exists\mathcal{O}\in\mathscr{O}(|\mathcal{D}|)$ s.t. $\mathcal{O}\subset \limsup \mathcal{D}$.
\end{enumerate}
\end{theorem}

Therefore, any two distinct beamforming vectors in an optimal d-codebook $\mathcal{D}$ with $|\mathcal{D}| \leq R$ are asymptotically orthogonal, and thus $\mathcal{D}$ converges asymptotically to an OMRS codebook. In particular, for codebooks with cardinality equal to $R$, Theorem \ref{sdtheorem} provides the same arguments as Theorem \ref{ldtheorem}. This follows from our previous observation that an OMRS codebook with cardinality equal to $R$ is also an SRS codebook.

From all the results that we have obtained up to now, we can conclude that OMRS is a universal necessary condition in the sense that for any SER-optimal d-codebook $\mathcal{D}$, there exists $\mathcal{O}\in\mathscr{O}(\min\{R,|\mathcal{D}|\})$ s.t. $\mathcal{O}\subset\limsup \mathcal{D}$. In other words, as $P$ grows to infinity, the upper limit of every optimal codebook should contain an OMRS codebook with the largest possible cardinality.
\section{Numerical Results}
\label{secNumerical}
In this section, we provide numerical evidence regarding the validity of our analytical results. For all the figures, the horizontal and the vertical axes represent $P$, and the SER, respectively.
\subsection{Diversity Bounds for Finite-Cardinality I-Codebooks}
\begin{figure}[h]
\begin{center}
\begin{psfrags}%
\psfragscanon%
%
\psfrag{s05}[t][t]{\color[rgb]{0,0,0}\setlength{\tabcolsep}{0pt}\begin{tabular}{c}$P$ (dB)\end{tabular}}%
\psfrag{s06}[b][b]{\color[rgb]{0,0,0}\setlength{\tabcolsep}{0pt}\begin{tabular}{c}SER\end{tabular}}%
\psfrag{s10}[][]{\color[rgb]{0,0,0}\setlength{\tabcolsep}{0pt}\begin{tabular}{c} \end{tabular}}%
\psfrag{s11}[][]{\color[rgb]{0,0,0}\setlength{\tabcolsep}{0pt}\begin{tabular}{c} \end{tabular}}%
\psfrag{s12}[l][l]{}%
\psfrag{s13}[l][l]{\color[rgb]{0,0,0}$\mathcal{C}_1$}%
\psfrag{s14}[l][l]{\color[rgb]{0,0,0}$\mathcal{C}_2$}%
\psfrag{s15}[l][l]{\color[rgb]{0,0,0}$\mathcal{C}_{\mathtt{SRS}} \cdot \mathbf{U}_2$}%
\psfrag{s16}[l][l]{\color[rgb]{0,0,0}$\mathcal{O}_1$}%
\psfrag{s17}[l][l]{\color[rgb]{0,0,0}$\mathcal{C}_3$}%
\psfrag{s18}[l][l]{\color[rgb]{0,0,0}$\mathcal{C}_{\mathtt{SRS}} \cdot \mathbf{U}_1$}%
\psfrag{s19}[l][l]{\color[rgb]{0,0,0}$\mathcal{C}_{\mathtt{SRS}}$}%
%
\psfrag{x01}[t][t]{$0$}%
\psfrag{x02}[t][t]{$10$}%
\psfrag{x03}[t][t]{$20$}%
\psfrag{x04}[t][t]{$30$}%
\psfrag{x05}[t][t]{$40$}%
\psfrag{x06}[t][t]{$50$}%
%
\psfrag{v01}[r][r]{$10^{-6}$}%
\psfrag{v02}[r][r]{$10^{-5}$}%
\psfrag{v03}[r][r]{$10^{-4}$}%
\psfrag{v04}[r][r]{$10^{-3}$}%
\psfrag{v05}[r][r]{$10^{-2}$}%
\psfrag{v06}[r][r]{$10^{-1}$}%
\psfrag{v07}[r][r]{$10^{0}$}%
%
\resizebox{13cm}{!}{\includegraphics{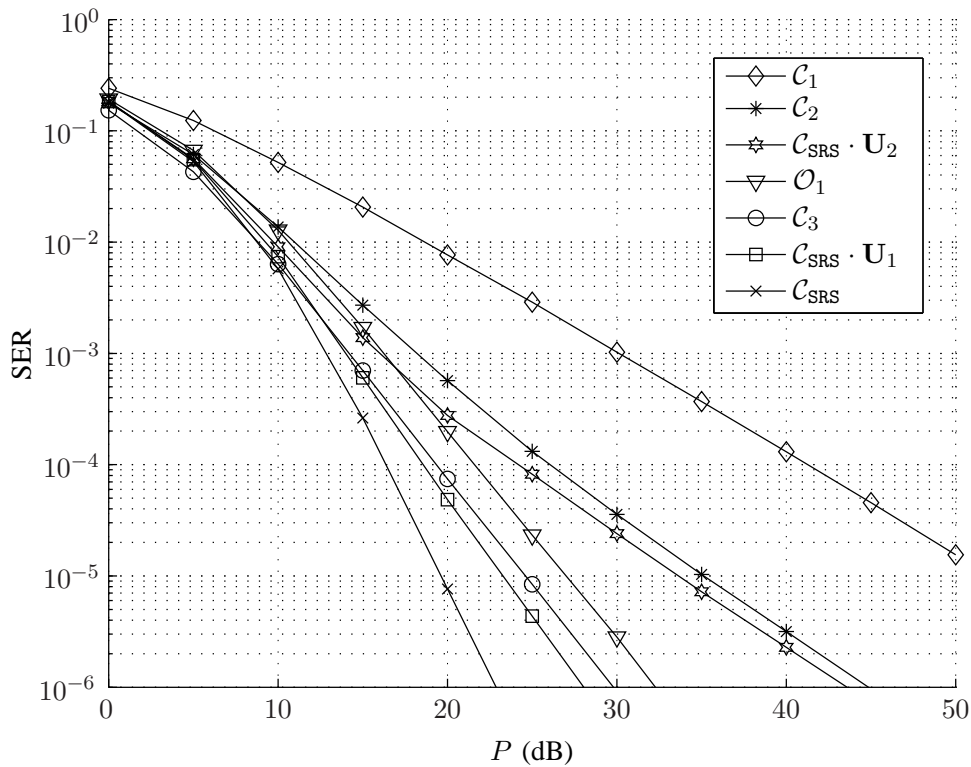}}%
\end{psfrags}
\end{center}
\caption{SERs with Different I-Codebooks.}
\label{letsimfig1}
\end{figure}

In Fig. \ref{letsimfig1}, we show the simulation results with i-codebooks for a $3$-relay network with power constraints $p_0 = 1$, $p_1 = 0.5$, $p_2 = p_3 = 2$, and channel variances $\sigma_{f_1}^2 = 1.2$, $\sigma_{f_2}^2 = 0.8$, $\sigma_{f_3}^2 = 1$, $\sigma_{g_1}^2 = 1.5$, $\sigma_{g_2}^2 = 1.7$, $\sigma_{g_3}^2 = 0.7$. The codebooks $\mathcal{C}_1$, $\mathcal{C}_2$, and $\mathcal{C}_3$ are as defined in (\ref{cbc1}), (\ref{cbc2}), and (\ref{cbc3}), respectively. $\mathcal{O}_1 = \{\,[1\,\,0\,\,0],\,[0\,-\!\!0.8\,\,1]\}$ is an OMRS codebook, and $\mathcal{C}_{\mathtt{SRS}}$ represents an arbitrary SRS codebook. $\mathcal{C}_{\mathtt{SRS}}\cdot\mathbf{U}_1$ and $\mathcal{C}_{\mathtt{SRS}}\cdot\mathbf{U}_2$ represent the transformations of an arbitrary SRS codebook by the unitary matrices $\mathbf{U}_1$ and $\mathbf{U}_2$ in (\ref{someunitaries}), respectively. Note that all SRS codebooks provide the same SER at any given $P$, as we have discussed in Section \ref{secProblemStatement} and as shown by Proposition \ref{snrinvariance}. Similarly, given any unitary matrix $\mathbf{U}$, all the codebooks $\mathcal{C}_{\mathtt{SRS}}\cdot\mathbf{U},\,\mathcal{C}_{\mathtt{SRS}} \in \mathscr{C}_{\mathtt{SRS}}$ provide the same SER at any given $P$.

%
%
%

We can observe from  Fig. \ref{letsimfig1} that $\diversity(\mathcal{C}_1) \leq 1$, $\diversity(\mathcal{C}_2) \leq 1$, and $\diversity(\mathcal{C}_3) \leq 2$, verifying Theorem \ref{sidetheorem}. Moreover, both codebooks seem to actually achieve their diversity bounds dictated by the theorem. This suggests that Theorem \ref{sidetheorem} also provides an accurate estimate on the diversity of any finite cardinality codebook. Also, $\mathcal{O}_1$ yields second order diversity as we have conjectured, and $\mathcal{C}_{\mathtt{SRS}}$ provides full-diversity.

We have analytically shown earlier in Example \ref{example3} that unlike a MISO system, in a relay network, (i) the performance of a codebook can significantly vary under unitary transformations, and (ii) the existence of linearly independent codebook vectors do not guarantee maximal diversity. Regarding the latter phenomenon, Fig. \ref{letsimfig1} demonstrates that even though $\mathcal{C}_2$ and $\mathcal{C}_3$ consist of $2$ and $3$ linearly independent codebook vectors, respectively, we have $\diversity(\mathcal{C}_2) \leq 1$ and $\diversity(\mathcal{C}_3) \leq 2$. For the former phenomenon, despite the fact that $\diversity(\mathcal{C}_{\mathtt{SRS}}) = 3$, we have $\diversity(\mathcal{C}_{\mathtt{SRS}} \cdot \mathbf{U}_2) \leq 1$ and $\diversity(\mathcal{C}_{\mathtt{SRS}} \cdot \mathbf{U}_1) \leq 2$, as we can infer from Fig. \ref{letsimfig1}. Hence, in relay networks, even the diversity provided by a codebook is not preserved under unitary transformations.

As a final remark for this set of simulations, we would like to note that we have chosen the power constraint scalers and channel variances in a random manner so as to demonstrate the validity of our results in ``asymmetric'' scenarios. We have obtained similar results for other (including identical) choices of these parameters.
\subsection{The Necessity of SRS}
\label{simthenecessofrs}
\begin{figure}[h]
\begin{center}
\begin{psfrags}%
\psfragscanon%
%
\psfrag{s09}[t][t]{\color[rgb]{0,0,0}\setlength{\tabcolsep}{0pt}\begin{tabular}{c}$P$ (dB)\end{tabular}}%
\psfrag{s10}[b][b]{\color[rgb]{0,0,0}\setlength{\tabcolsep}{0pt}\begin{tabular}{c}SER\end{tabular}}%
\psfrag{s14}[][]{\color[rgb]{0,0,0}\setlength{\tabcolsep}{0pt}\begin{tabular}{c} \end{tabular}}%
\psfrag{s15}[][]{\color[rgb]{0,0,0}\setlength{\tabcolsep}{0pt}\begin{tabular}{c} \end{tabular}}%
\psfrag{s16}[l][l]{\color[rgb]{0,0,0}}%
\psfrag{s17}[l][l]{\color[rgb]{0,0,0}$\widetilde{\mathcal{C}}(1,1),\,\widetilde{\mathcal{C}}(1,2)$}%
\psfrag{s18}[l][l]{\color[rgb]{0,0,0}$\widetilde{\mathcal{C}}(\frac{1}{4},1),\,\widetilde{\mathcal{C}}(\frac{1}{4},2)$}%
\psfrag{s19}[l][l]{\color[rgb]{0,0,0}$\widetilde{\mathcal{D}}_1,\,\widetilde{\mathcal{D}}_2$}%
\psfrag{s20}[l][l]{\color[rgb]{0,0,0}$\widetilde{\mathcal{C}}(\frac{1}{16},1),\,\widetilde{\mathcal{C}}(\frac{1}{16},\!2)$}%
\psfrag{s21}[l][l]{\color[rgb]{0,0,0}$\mathcal{C}_{\mathtt{SRS}}$}%
\psfrag{s22}[l][l]{\color[rgb]{0,0,0}$\mathcal{X}$}%
%
\psfrag{x01}[t][t]{0}%
\psfrag{x02}[t][t]{0.1}%
\psfrag{x03}[t][t]{0.2}%
\psfrag{x04}[t][t]{0.3}%
\psfrag{x05}[t][t]{0.4}%
\psfrag{x06}[t][t]{0.5}%
\psfrag{x07}[t][t]{0.6}%
\psfrag{x08}[t][t]{0.7}%
\psfrag{x09}[t][t]{0.8}%
\psfrag{x10}[t][t]{0.9}%
\psfrag{x11}[t][t]{1}%
\psfrag{x12}[t][t]{10}%
\psfrag{x13}[t][t]{15}%
\psfrag{x14}[t][t]{20}%
\psfrag{x15}[t][t]{25}%
\psfrag{x16}[t][t]{30}%
\psfrag{x17}[t][t]{35}%
%
\psfrag{v01}[r][r]{0}%
\psfrag{v02}[r][r]{0.1}%
\psfrag{v03}[r][r]{0.2}%
\psfrag{v04}[r][r]{0.3}%
\psfrag{v05}[r][r]{0.4}%
\psfrag{v06}[r][r]{0.5}%
\psfrag{v07}[r][r]{0.6}%
\psfrag{v08}[r][r]{0.7}%
\psfrag{v09}[r][r]{0.8}%
\psfrag{v10}[r][r]{0.9}%
\psfrag{v11}[r][r]{1}%
\psfrag{v12}[r][r]{$10^{-7}$}%
\psfrag{v13}[r][r]{$10^{-6}$}%
\psfrag{v14}[r][r]{$10^{-5}$}%
\psfrag{v15}[r][r]{$10^{-4}$}%
\psfrag{v16}[r][r]{$10^{-3}$}%
\psfrag{v17}[r][r]{$10^{-2}$}%
%
\resizebox{13cm}{!}{\includegraphics{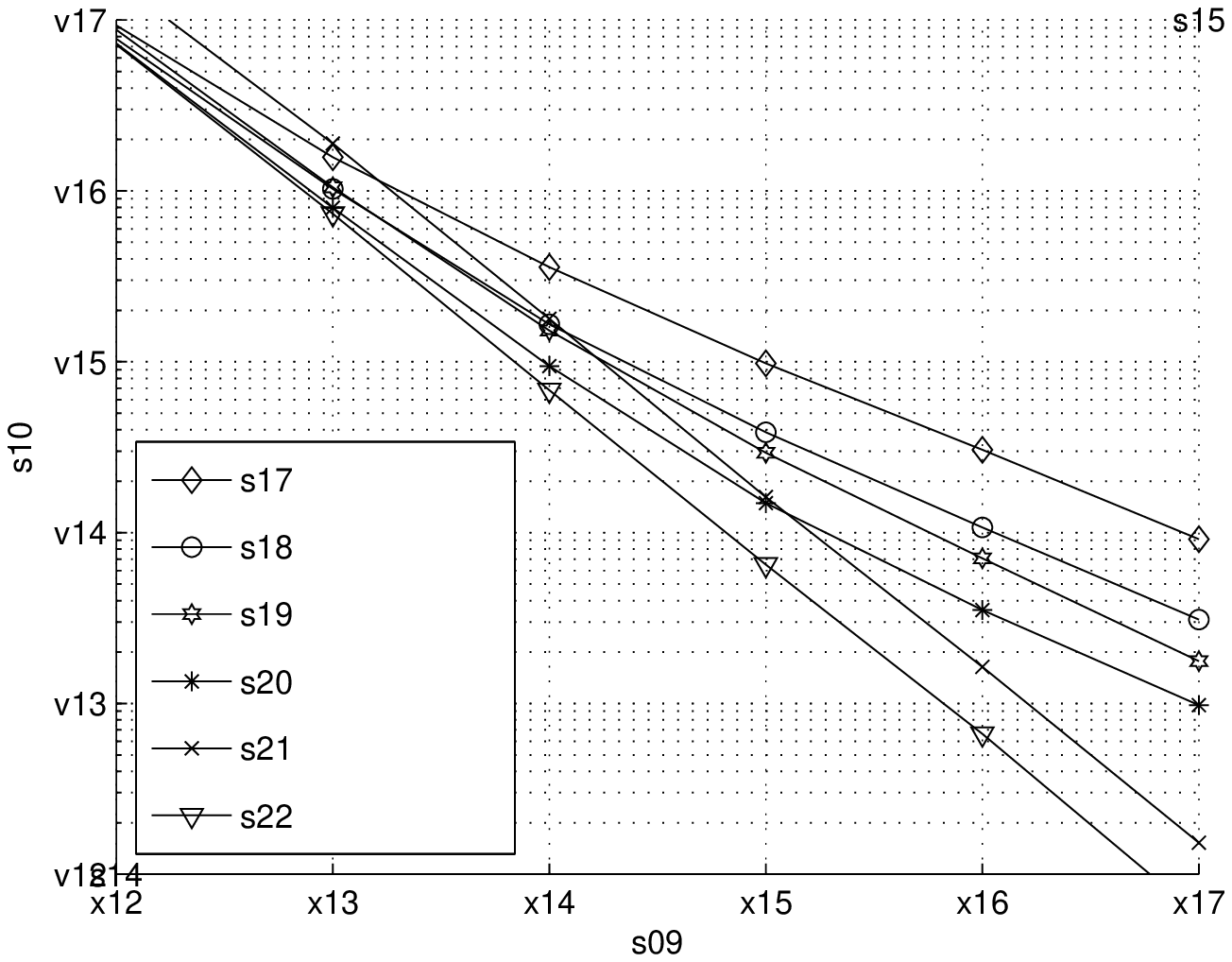}}%
\end{psfrags}%
\end{center}
\caption{An illustration of the validity of Theorems \ref{litheorem} and \ref{ldtheorem} for $R=2$.}
\label{letsimfig2}
\end{figure}

Let us now demonstrate the validity of Theorems \ref{litheorem} and \ref{ldtheorem} for a network with $R=2$. We assume that the power constraint scalers and the channel variances of the network are equal to unity. In this set of simulations, we use a special type of codebook that we define in what follows: For any $0 \leq \epsilon \leq 1$, let
\begin{align}
\label{specialcodebookdefi}
\widetilde{\mathcal{C}}(\epsilon, r) & \triangleq \{\mathbf{x}:\mathbf{x}\in\mathcal{X},\,|x_r|^2 \geq \epsilon\}.
\end{align}
 In Fig. \ref{letsimfig2}, we show the SERs for our $2$-relay network with i-codebooks $\widetilde{\mathcal{C}}(\epsilon, r),\,\epsilon = 1,\,\frac{1}{4},\,\frac{1}{16},\,r=1,2$, $\mathcal{C}_{\mathtt{SRS}}$, $\mathcal{X}$, and the d-codebooks $\smash{\widetilde{\mathcal{D}}}_r \triangleq \widetilde{\mathcal{C}}(\frac{1}{\log P}, r),\,r=1,2$. Note that, as a result of our choice of the network parameters, the SER with $\smash{\widetilde{\mathcal{C}}}(\epsilon,1)$ is the same as the SER with $\smash{\widetilde{\mathcal{C}}}(\epsilon,2)$ at any given $P$. Similarly, the SER with $\smash{\widetilde{\mathcal{D}}}_1$ is the same as the SER with $\smash{\widetilde{\mathcal{D}}}_2$ at any given $P$.

%
%
%

We first demonstrate the validity of Theorem \ref{litheorem}. Since $\{r\}\in\mathscr{R}(\smash{\widetilde{\mathcal{C}}}({\epsilon,r}))$, and $\xi(\smash{\widetilde{\mathcal{C}}}(\epsilon,r), \{r\}) = \epsilon$, by Lemma \ref{mainlemma}, $\smash{\widetilde{\mathcal{C}}}(\epsilon,r)$ provides at most a diversity of $1$ for any fixed $\epsilon>0$. This is precisely what we observe in Fig. \ref{letsimfig2}. In general, we expect a similar behavior for any given $\epsilon > 0$. Thus, if we use an i-codebook $\mathcal{C}$ with either $\mathcal{C}\subset\smash{\widetilde{\mathcal{C}}}(\epsilon, 1)$ or $\mathcal{C}\subset\smash{\widetilde{\mathcal{C}}}(\epsilon, 2)$ for some $\epsilon > 0$, $\mathcal{C}$ will not be able to provide diversity more than $1$. In other words, if $\mathcal{C}\subset\bigcup_{\epsilon > 0} \widetilde{\mathcal{C}}(\epsilon, 1)$ or $\mathcal{C}\subset\bigcup_{\epsilon > 0} \widetilde{\mathcal{C}}(\epsilon, 2)$, then $\diversity(\mathcal{C}) \leq 1$. Hence, if $\mathcal{C}^{\star}$ is an i-codebook that achieves diversity $2$, then $\exists\mathbf{e}_1^{\star},\mathbf{e}_2^{\star}\in\mathcal{C}^{\star}$ s.t.
\begin{align}
\mathbf{e}_1^{\star} \in \left[ \bigcup_{\epsilon > 0} \widetilde{\mathcal{C}}(\epsilon, 1) \right]^c = \bigcap_{\epsilon > 0} \left[ \widetilde{\mathcal{C}}(\epsilon, 1) \right]^c = \bigcap_{\epsilon > 0}\{\mathbf{x}\in\mathcal{X}:|x_1|^2 < \epsilon\}
 = \{\mathbf{x}\in\mathcal{X}:|x_1|=0\},
\end{align}
and $\mathbf{e}_2^{\star} \in \{\mathbf{x}\in\mathcal{X}:|x_2|=0\}$, where $\mathcal{C}^c \triangleq \mathcal{X} - \mathcal{C}$. Note that $\mathbf{e}_1^{\star}$ and $\mathbf{e}_2^{\star}$ are SRS vectors. Therefore, if $\mathcal{C}^{\star}$ achieves full diversity, it should contain an SRS codebook. This verifies Theorem \ref{litheorem}.

The verification of Theorem \ref{ldtheorem} is analogous:  Let $\mathcal{D}^\star$ denote an optimal d-codebook, and $\epsilon_P \triangleq \sup \{\epsilon: \mathcal{D}_P^{\star} \subset \smash{\widetilde{\mathcal{C}}}(\epsilon, 1)\mbox{ or } \mathcal{D}_P^{\star} \subset \smash{\widetilde{\mathcal{C}}}(\epsilon, 2)\}$. Since $\mathcal{D}^\star$ is an optimal d-codebook, it achieves full-diversity. Thus, using the same arguments in the previous paragraph, $\epsilon_P\rightarrow 0$ as $P\rightarrow\infty$. On the other hand, by the definition of $\epsilon_P$, we have $\exists\mathbf{x}_{r,P}^{\star} \in \mathcal{D}_P^{\star}$ s.t. $|x_{r,P,r}^{\star}|^2 \leq \epsilon_P + \epsilon_{r,P}',\,r=1,2$, where $\epsilon_{r,P}' > 0$ can be chosen arbitrarily. Let us choose $\epsilon_{r,P}' = \epsilon_P,\,r=1,2$. Then, we have $|x_{r,P,r}^{\star}|^2 \leq 2\epsilon_P,\,r=1,2$. This shows the existence of two beamforming vectors in $\mathcal{D}_P^{\star}$, namely $\mathbf{x}_{1,P}^{\star}$ and $\mathbf{x}_{2,P}^{\star}$, that converges to two distinct SRS vectors as $P\rightarrow\infty$. This verifies the limit arguments in Theorem \ref{ldtheorem}.

Theorem \ref{ldtheorem} also provides an estimate on how fast $\epsilon_P$ should decay. The performance of the d-codebooks $\widetilde{\mathcal{D}}_1$ and $\widetilde{\mathcal{D}}_2$ in Fig. \ref{letsimfig2} demonstrate that the decay should be no slower than $\tfrac{1}{\log P}$, and thus verifies the theorem. On the other hand, since both codebooks do not provide maximal diversity, the estimate of Theorem \ref{ldtheorem} might be rather loose.
\subsection{The Necessity of OMRS}
We now demonstrate the validity of Theorems \ref{sitheorem} and \ref{sdtheorem} for a network with $R=3$. We assume that the power constraint scalers and the channel variances of the network are equal to unity. Our goal is to determine the structure of optimal codebooks that have cardinality equal to $2$ and thus provide a diversity of $2$. For that purpose, similar to what we have done in Section \ref{simthenecessofrs}, we use the special i-codebook $\widetilde{\mathcal{C}}(\epsilon, r)$ as defined in (\ref{specialcodebookdefi}).

\begin{figure}[h]
\begin{center}
\begin{psfrags}%
\psfragscanon%
%
\psfrag{s05}[t][t]{\color[rgb]{0,0,0}\setlength{\tabcolsep}{0pt}\begin{tabular}{c}P (dB)\end{tabular}}%
\psfrag{s06}[b][b]{\color[rgb]{0,0,0}\setlength{\tabcolsep}{0pt}\begin{tabular}{c}SER\end{tabular}}%
\psfrag{s10}[][]{\color[rgb]{0,0,0}\setlength{\tabcolsep}{0pt}\begin{tabular}{c} \end{tabular}}%
\psfrag{s11}[][]{\color[rgb]{0,0,0}\setlength{\tabcolsep}{0pt}\begin{tabular}{c} \end{tabular}}%

\psfrag{s12}[l][l]{\color[rgb]{0,0,0}}%
\psfrag{s13}[l][l]{\color[rgb]{0,0,0}$\widetilde{\mathcal{C}}(1,1),\,\widetilde{\mathcal{C}}(1,2),\,\widetilde{\mathcal{C}}(1,3)$}%
\psfrag{s14}[l][l]{\color[rgb]{0,0,0}$\widetilde{\mathcal{C}}(\frac{1}{4},1),\,
\widetilde{\mathcal{C}}(\frac{1}{4},2),\,\widetilde{\mathcal{C}}(\frac{1}{4},3)$}%
\psfrag{s15}[l][l]{\color[rgb]{0,0,0}$\widetilde{\mathcal{D}}_1,\,\widetilde{\mathcal{D}}_2,\,\widetilde{\mathcal{D}}_3$}%
\psfrag{s16}[l][l]{\color[rgb]{0,0,0}$\widetilde{\mathcal{C}}(\frac{1}{16},1),\,
\widetilde{\mathcal{C}}(\frac{1}{16},2),\,\widetilde{\mathcal{C}}(\frac{1}{16},3)$}%
\psfrag{s17}[l][l]{\color[rgb]{0,0,0}$\mathcal{C}_{\mathtt{SRS}}$}%
\psfrag{s18}[l][l]{\color[rgb]{0,0,0}$\mathcal{X}$}%
%
\psfrag{x01}[t][t]{0}%
\psfrag{x02}[t][t]{0.1}%
\psfrag{x03}[t][t]{0.2}%
\psfrag{x04}[t][t]{0.3}%
\psfrag{x05}[t][t]{0.4}%
\psfrag{x06}[t][t]{0.5}%
\psfrag{x07}[t][t]{0.6}%
\psfrag{x08}[t][t]{0.7}%
\psfrag{x09}[t][t]{0.8}%
\psfrag{x10}[t][t]{0.9}%
\psfrag{x11}[t][t]{1}%
\psfrag{x12}[t][t]{5}%
\psfrag{x13}[t][t]{10}%
\psfrag{x14}[t][t]{15}%
\psfrag{x15}[t][t]{20}%
\psfrag{x16}[t][t]{25}%
%
\psfrag{v01}[r][r]{0}%
\psfrag{v02}[r][r]{0.1}%
\psfrag{v03}[r][r]{0.2}%
\psfrag{v04}[r][r]{0.3}%
\psfrag{v05}[r][r]{0.4}%
\psfrag{v06}[r][r]{0.5}%
\psfrag{v07}[r][r]{0.6}%
\psfrag{v08}[r][r]{0.7}%
\psfrag{v09}[r][r]{0.8}%
\psfrag{v10}[r][r]{0.9}%
\psfrag{v11}[r][r]{1}%
\psfrag{v12}[r][r]{$10^{-7}$}%
\psfrag{v13}[r][r]{$10^{-6}$}%
\psfrag{v14}[r][r]{$10^{-5}$}%
\psfrag{v15}[r][r]{$10^{-4}$}%
\psfrag{v16}[r][r]{$10^{-3}$}%
\psfrag{v17}[r][r]{$10^{-2}$}%
%
\resizebox{13cm}{!}{\includegraphics{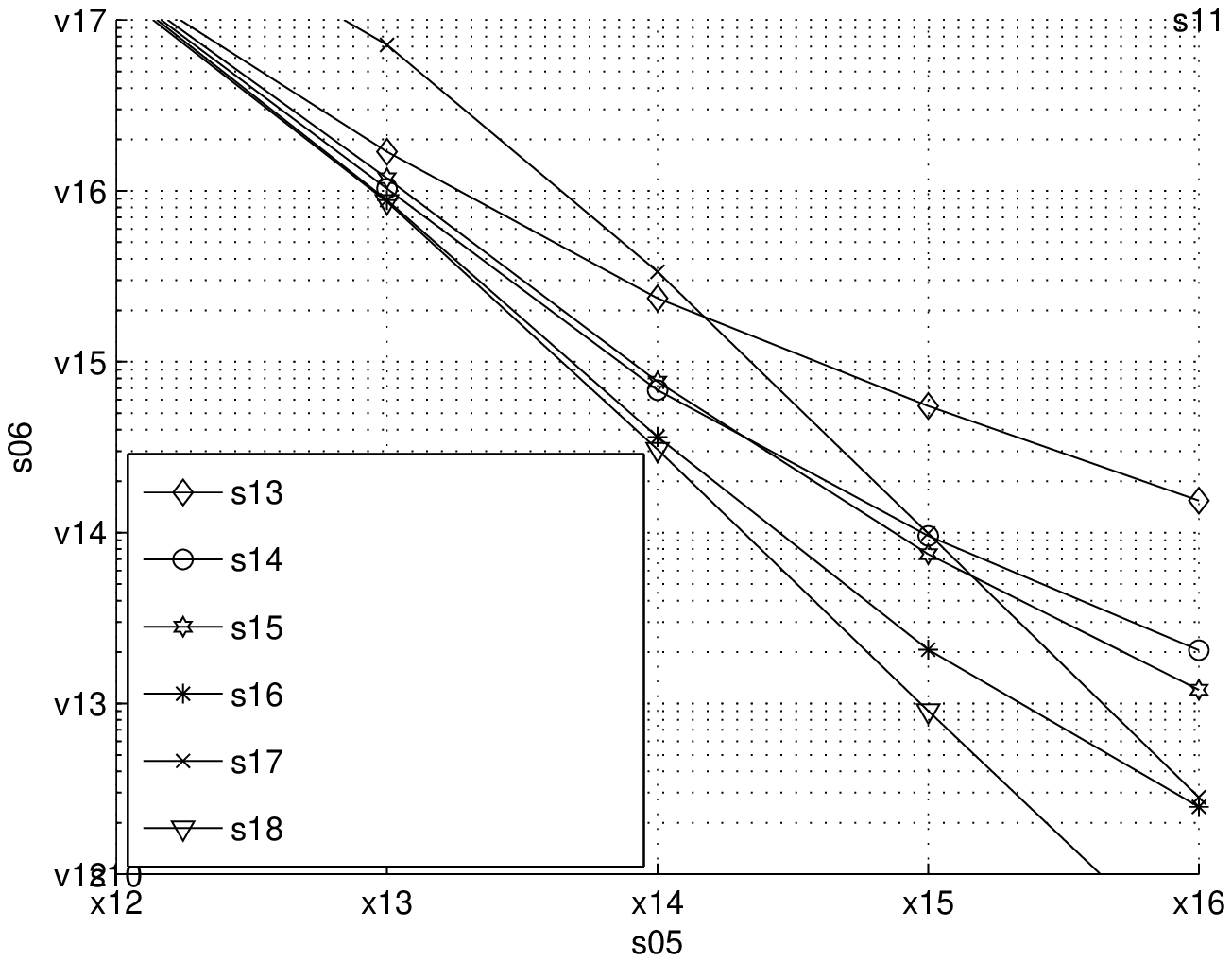}}%
\end{psfrags}%
\end{center}
\caption{An illustration of the validity of Theorems \ref{sitheorem} and \ref{sdtheorem} for $R=3$.}
\label{letsimfig3}
\end{figure}

In Fig. \ref{letsimfig3}, we show the SERs for our $3$-relay network with i-codebooks $\widetilde{\mathcal{C}}(\epsilon, r),\,\epsilon = 1,\,\frac{1}{4},\,\frac{1}{16},\,r=1,2,3$, $\mathcal{C}_{\mathtt{SRS}}$, $\mathcal{X}$, and the d-codebooks $\smash{\widetilde{\mathcal{D}}}_r \triangleq \widetilde{\mathcal{C}}(\frac{1}{\log P}, r),\,r=1,2,3$. As a result of our choice of the network parameters, for a given $\epsilon$, the SERs with $\smash{\widetilde{\mathcal{C}}}(\epsilon,1),\,\smash{\widetilde{\mathcal{C}}}(\epsilon,2)$ and $\smash{\widetilde{\mathcal{C}}}(\epsilon,3)$ are the same at any given $P$. Similarly, the SERs with $\smash{\widetilde{\mathcal{D}}}_1,\,\smash{\widetilde{\mathcal{D}}}_1$ and $\smash{\widetilde{\mathcal{D}}}_3$ are the same at any given $P$.

%
%
%

We first demonstrate the validity of Theorem \ref{litheorem}. We can observe from Fig. \ref{letsimfig2} that $\smash{\widetilde{\mathcal{C}}}(\epsilon,r)$ provides at most a diversity of $1$ for any fixed $\epsilon>0$. In general, we expect a similar behavior for any given $\epsilon > 0$. Thus, if we use an i-codebook $\mathcal{C}$ with  $\mathcal{C}\subset\smash{\widetilde{\mathcal{C}}}(\epsilon, r)$ for some $r\in\{1,2,3\}$ and $\epsilon > 0$, then $\diversity(\mathcal{C}) \leq 1$.
As a result, using the same arguments in Section \ref{simthenecessofrs}, if $\diversity(\mathcal{C}^*) = 2$, then 
$\exists\mathbf{y}_r\in\mathcal{C}^*\mbox{ s.t. }\mathbf{y}_r^* \in \{\mathbf{y}\in\mathcal{X}:|y_r|=0\},\,r=1,2,3$.
In other words, for any $r\in\{1,2,3\}$, there exists a beamforming vector in $\mathcal{C}^* = \{\mathbf{x}_1^*, \mathbf{x}_2^*\}$ with a vanishing $r$th component. Therefore, $\sum_{r=1}^3 |x_{1r}^*| |x_{2r}^*| = 0$, which means that $\mathcal{C}^*$ is an OMRS codebook. This verifies Theorem \ref{sitheorem}.

 In order to verify Theorem \ref{sdtheorem}, let $\mathcal{D}^\star$ with $\mathcal{D}_P^* = \{\mathbf{x}_{1,P}^*,\mathbf{x}_{2,P}^*\},\,P\in\mathbb{R}$ and $|\mathcal{D}^{\star}| = 2$ denote an optimal small d-codebook, and $\epsilon_P \triangleq \sup \{\epsilon: \exists r\in\{1,2,3\}\mbox{ s.t. }\mathcal{D}_P^{\star} \subset \smash{\widetilde{\mathcal{C}}}(\epsilon, r)\}$. Since $\mathcal{D}^\star$ is optimal, it achieves second order diversity. Using the same arguments in the previous paragraph, $\epsilon_P\rightarrow 0$ as $P\rightarrow\infty$. On the other hand, $\exists\mathbf{y}_{r,P}^{\star} \in \mathcal{D}_P^{\star}$ s.t. $|y_{r,P,r}^{\star}| \leq \sqrt{2\epsilon_P},\,r=1,2,3,$ by the definition of $\epsilon_P$. As a result, $\sum_{r=1}^3 |x_{1,r,P}^*| |x_{2,r,P}^*| \leq 6\epsilon_P,\,\forall P\in\mathbb{R}$, and $\sum_{r=1}^3 |x_{1,r,P}^*| |x_{2,r,P}^*| \rightarrow 0$ as $P\rightarrow\infty$. In other words, the two beamforming vectors in $\mathcal{D}_P^*$ should become asymptotically orthogonal. Finally, the performance of the codebooks $\widetilde{\mathcal{D}}_r,\,r=1,2,3$ in Fig. \ref{letsimfig3} demonstrate that $\epsilon_P$ should decay no slower than $\frac{1}{\log P}$. These verify Theorem \ref{sdtheorem}.

\section{Conclusions}
\label{secConc}
We have determined some necessary structural properties of symbol error rate optimal quantizers for limited feedback beamforming in wireless networks with a single transmitter-receiver pair and $R$ parallel amplify-and-forward relays. We have shown that any power-independent codebook (i-codebook) necessarily contains an orthogonal multiple-relay selection (OMRS) codebook with the largest possible cardinality. In particular, if the cardinality of the codebook is no less than $R$, an i-codebook achieves maximal diversity if and only if it contains the single-relay selection (SRS) codebook. We have obtained similar results for the general case of power-dependent codebooks (d-codebooks): An optimal d-codebook should contain an OMRS codebook with the largest possible cardinality, asymptotically as the transmitter powers grow to infinity.
\appendices
\section{Proof of Lemma \ref{mainlemma}}
\label{proofofmainlemma}
Note that $\mathcal{R}\neq\emptyset$, since if $\mathcal{R}=\emptyset$ then $\mathbf{x} = \mathbf{0},\,\forall\mathbf{x}\in\mathcal{C}$, contradicting Proposition \ref{codebookoptimalitylemma}.

Using (\ref{receivedsnr}), for any set of indices $\mathcal{R}\neq\emptyset$ and relay normalization factors $\rho_r'\leq\rho_r,\,\forall r$, the SNR with any beamforming vector $\mathbf{x}\in\mathcal{C}$, can be upper bounded by
\begin{align}
  \mathtt{SNR}_P(\mathbf{x}, \mathbf{h})  & \leq \frac{RP_0 \sum_{r=1}^R |x_rf_rg_r|^2 \rho_r'}{1 + \sum_{r=1}^R |x_rg_r|^2\rho_r'} \\
   & \leq \frac{RP_0 \sum_{r=1}^R |x_rf_rg_r|^2 \rho_r}{1 + \sum_{r=1}^R |x_rg_r|^2\rho_r} \\ \label{longinek1} & = \frac{R\sum_{r=1}^R \frac{|x_r|^2|f_{r}|^2P_0 |g_{r}|^2 P_r}{1+|f_r|^2 P_0}}{1 + \sum_{r=1}^R \frac{|x_r|^2|g_{r}|^2  P_r}{1+|f_r|^2P_0}},
 \end{align}
 where the first and the second inequalities follow from H\"{o}lder's inequality, and the fact that $\rho_r' \leq \rho_r,\,\forall r$, respectively. The proof of the lemma for $R=1$ is now straightforward. If $R=1$, we have $|x_1| = 1,\,\forall \mathbf{x}\in\mathcal{C}$, and thus $\xi(\mathcal{C},\mathcal{R}) = 1$ for any $\mathcal{C}$ and $\mathcal{R}$ (indeed the only available $\mathcal{R}$ will be $\mathcal{R} = \{1\}$). Using (\ref{longinek1}), we have $\mathtt{SNR}_P(\mathbf{x}, \mathbf{h})  \leq \frac{|f_1|^2P_0 |g_1|^2P_1}{1 + |f_1|^2P_0 +  |g_1|^2P_1} \leq |f_1|^2P_0$. Since this final upper bound is the SNR of a fading channel with single transmitter and receiver antennas, we have $\mathtt{SER}_P(\mathcal{Q}^{\star}_{P, \mathcal{C}}) \geq \letcpref P^{-1}$ for some constant $0<\letcpref<\infty$ independent of $P$.

 For $R \geq 2$, we shall further bound $\mathtt{SNR}_P(\mathbf{x}, \mathbf{h})$. For the numerator of (\ref{longinek1}), we have
 \begin{multline}
  \label{numbound}
 R \sum_{r=1}^R \frac{|x_r|^2|f_{r}|^2P_0|g_{r}|^2 P_r}{1+|f_r|^2 P_0} \leq  R \sum_{r=1}^R \frac{|f_{r}|^2P_0|g_{r}|^2 P_r}{1+|f_r|^2 P_0} \leq R \sum_{r=1}^R |g_{r}|^2 P_r \leq R^2 \max_r |g_r|^2 P_r \\ = R^2 P \max_r \{\sigma_{g_r}^{2}\sigma_{g_r}^{-2}|g_r|^2 p_r\} \leq R^2 \max_r \{p_r\sigma_{g_r}^2\}  P \max_r Z_r.
 \end{multline}
 where $Z_r \triangleq \sigma_{g_r}^{-2}|g_r|^2$. Note that $Z_r \sim \Gamma(1,1)$.
 Now, for the denominator of (\ref{longinek1}),
 \begin{multline}
 \label{denombound}
 1 + \sum_{r=1}^R \frac{|x_r|^2|g_{r}|^2  P_r}{1+|f_r|^2P_0} \geq \sum_{r=1}^R \frac{|x_r|^2|g_{r}|^2  P_r}{1+|f_r|^2P_0} \geq \max_r|x_r|^2 \min_r \frac{|g_{r}|^2  P_r}{1+|f_r|^2P_0} \\ \geq  \max_r|x_r|^2 \frac{\min_{r\in\mathcal{R}} |g_r|^2 P_r}{\max_{r\in\mathcal{R}} (1+|f_r|^2 P_0)} \geq  \frac{\max_r|x_r|^2 \min_r \{p_r\sigma_{g_r}^2\} P \min_r Z_r}{\max\{1, p_0\max_r\sigma_{f_r}^2\}\left(1 + P \sum_{r\in\mathcal{R}} \sigma_{f_r}^{-2}|f_r|^2\right)}.
 \end{multline}
 Now let $V \triangleq YZ$ with $Y \triangleq \frac{1}{P} + \sum_{r\in\mathcal{R}}\sigma_{f_r}^{-2}|f_{r}|^2 $ and $Z \triangleq \frac{\max_r Z_r }{\min_r Z_r}$. Using (\ref{denombound}) and (\ref{numbound}) in the final upper bound in (\ref{longinek1}), and then taking the supremum over all possible $\mathbf{x}\in\mathcal{C}$, we have
\begin{align}
\label{snruboundofaquantizer}
 \mathtt{SNR}_P(\mathcal{Q}^{\star}_{P, \mathcal{C}}(\mathbf{h}), \mathbf{h}) &   \leq \letcf PV,
\end{align}
where $\letcf \triangleq [\xi(\mathcal{C},\mathcal{R})]^{-1}\letcg$ is a finite constant with $\xi(\mathcal{C}, \mathcal{R})$ is as defined in the statement of the lemma, and $\letcg \triangleq R^2 \max\{1,p_0\max_r\sigma_{f_r}^2\}\max_r \{p_r\sigma_{g_r}^2\}/ \min_r \{p_r\sigma_{g_r}^2\}$. Note that since $\letcg \geq 1$ and $\xi(\mathcal{C}, \mathcal{R}) \leq 1$, we have $\letcf \geq 1$.

Now, substituting (\ref{snruboundofaquantizer}) to (\ref{serofaquantizer}), we have
\begin{align}
\label{mainlbound}
\mathtt{SER}_P(\mathcal{Q}^{\star}_{P, \mathcal{C}}) &  \geq   \int_{0}^{\infty} \mathrm{Q}(\sqrt{2\letcf vP})f(v)\mathrm{d}v.
\end{align}
In the following, we find a lower bound on the PDF of $V=YZ$. Since $Y$ is the sum of $R$ independent $\Gamma(1,1)$ random variables and a constant $\frac{1}{P}$, it follows a ``shifted'' gamma distribution:
\begin{align}
 \label{ybond}
  f_Y(y) = \frac{1}{\Gamma(|\mathcal{R}|)}e^{-\left(y-\frac{1}{P}\right)}
\left(y-\frac{1}{P}\right)^{|\mathcal{R}|-1},\,y \geq \frac{1}{P}.
\end{align}
Now, let us evaluate $f_Z(z)$. Note that for $z<1$, $F_Z(z) = 0$, and thus $f_Z(z) = 0,\,z<1$. For $z\geq 1$, the CDF of $Z$ can be expressed as $F_Z(z) = P(E)$ where $E$ is the event that $\max_r Z_r \leq z \min_r Z_r$, with $Z_r \triangleq |g_r|^2$. Moreover, $E$ is the union of $R(R-1)$ disjoint events $E_{ij},\,i\neq j,\,i,j\in\{1,\ldots, R\}$, where $E_{ij}$ is the event that $Z_i = \min Z_r,\,Z_j = \max Z_r,\, Z_i\in[0,\infty),\,Z_j\in(Z_i,zZ_i],\,Z_k\in(Z_i,Z_j),\,k\neq i,\,k\neq j$.\footnote{We ignore the events that have zero probability, e.g. the event that $\max Z_r = \min Z_r$} Since $Z_r$ are identically distributed, and each $E_{ij}$ has the same probability, for $z\geq 1$, we have
\begin{align}
\label{aciklaeq1} F_Z (z) &  =   R(R\!-\!1) \int_0^{\infty} \int_x^{xz} \underbrace{\int_x^y \cdots \int_x^y}_{R-2\,\,\mathrm{integrals}} e^{-x-y-\sum_{i}\!\!w_i} \prod_{i} \mathrm{d}w_i \mathrm{d}y \mathrm{d}x \\
\label{sonradan1} &   =   R(R\!-\!1) \int_0^{\infty} \int_x^{xz}  e^{-x-y} (e^{-x} - e^{-y})^{R-2}\mathrm{d}y \mathrm{d}x \\
\label{sonradan2} &   =   R(R\!-\!1) \int_0^{\infty} \int_x^{xz}  e^{-x-y} \sum_{r=0}^{R-2} {R-2 \choose r} (-1)^r e^{-yr}  e^{-x(R-2-r)}\mathrm{d}y \mathrm{d}x\\
\label{aciklaeq2} &   = (R-1)\sum_{r=0}^{R-2}{R-2 \choose r} \frac{(-1)^r(z-1)}{R+(z-1)(1+r)}\\
\label{aciklaeq3} &   = (z\!-\!1)(R\!-\!1)\int_0^{\infty}e^{-x(R+(z-1))} \sum_{r=0}^{R-2}{R-2 \choose r}(-1)^r e^{-rx(z-1)} \mathrm{d}x \\
\label{aciklaeq4} &   = (z\!-\!1)(R\!-\!1)\int_0^{\infty}\!\!e^{-x(R+(z-1))} (1\! - \! e^{-x(z-1)})^{R-2} \mathrm{d}x \\
\label{aciklaeq5} &   = 2^{R-2}(z\!-\!1)(R\!-\!1)\int_0^{\infty}\!\!e^{-x\frac{R(z+1)}{2}} \sinh^{R-2}\left[\frac{x(z-1)}{2}\right] \mathrm{d}x \\
\label{aciklaeq6} &   = \frac{\Gamma(R)\Gamma(1+\frac{R}{z-1})}{\Gamma(R+\frac{R}{z-1})} \\
\label{aciklaeq7} & = \frac{\Gamma(R)}{ \prod_{r=1}^{R-1} (r + \frac{R}{z-1})},
\end{align}
where (\ref{sonradan2}) follows from the binomial expansion of the term $(e^{-x} - e^{-y})^{R-2}$ in (\ref{sonradan1}). In order to obtain (\ref{aciklaeq3}), we have rewritten the denominator of the fraction in (\ref{aciklaeq2}) in integral form by using the identity $\int_0^{\infty}e^{-\alpha x}\mathrm{d}x = \frac{1}{\alpha},\,\alpha > 0$. Also, (\ref{aciklaeq4}) is a result of the fact that $\sum_{r=0}^{R-2} (-1)^r \beta^r = (1-\beta)^{R-1}$ for $0\leq\beta\leq 1$, and (\ref{aciklaeq6}) follows from \cite[Eq. 3.541.1]{toi}. In order to derive (\ref{aciklaeq7}) from (\ref{aciklaeq6}), we have used the identity $\Gamma(1+x) = x\Gamma(x),\,x\in\mathbb{R}$, which implies $\Gamma(R+\frac{R}{z-1}) = (R-1+\frac{R}{z-1})\Gamma(R-1+\frac{R}{z-1}) = \prod_{r=1}^{R-1} (r+\frac{R}{z-1}) \Gamma(1 + \frac{R}{z-1})$.

We can now find the PDF of $Z$ using (\ref{aciklaeq7}). We have
\begin{align}
f_Z(z) & =   \frac{\partial}{\partial z}F_Z(z) \\
& = \frac{-\Gamma(R)}{ \prod_{r=1}^{R-1} (r + \frac{R}{z-1})^2}\frac{\partial}{\partial z}\left\{\prod_{r=1}^{R-1} \left(r + \frac{R}{z-1}\right)  \right\} \\
& = \frac{-\Gamma(R)}{ \prod_{r=1}^{R-1} (r + \frac{R}{z-1})^2}\sum_{r=1}^R\frac{\partial}{\partial z}\left\{r + \frac{R}{z-1}\right\}\prod_{\substack{q=1 \\ q\neq r}}^{R-1}\left(q + \frac{R}{z-1}\right) \\
& = \frac{-\Gamma(R)}{ \prod_{r=1}^{R-1} (r + \frac{R}{z-1})^2}\sum_{r=1}^{R-1}\frac{-R}{(z-1)^2}\frac{\prod_{q=1}^{R-1}(q + \frac{R}{z-1})}{r + \frac{R}{z-1}} \\
& = \frac{\Gamma(R+1)\sum_{r=1}^{R-1}(r+\frac{R}{z-1})^{-1}}{(z-1)^2\prod_{r=1}^{R-1}(r+\frac{R}{z-1})} \\ & \geq \frac{\Gamma(R+1)\sum_{r=1}^{R-1}(R+\frac{R}{z-1})^{-1}}{(z-1)^2\prod_{r=1}^{R-1}(R+\frac{R}{z-1})}  \\ \label{zbond} & =    \frac{(R-1)\Gamma(R+1)}{R^R} \frac{(z-1)^{R-2}}{z^R}.
\end{align}
Now, we find a lower bound on the PDF of $V = YZ$. It can be shown\cite{aofrvs} that the PDF of $V$ is given by $f_V(v) = \int_{-\infty}^{\infty} f_{Z}(x) f_{Y}\left(\frac{v}{x}\right)\frac{1}{|x|}\mathrm{d}x$. Substituting the PDF of $Y$ in (\ref{ybond}), and the lower bound in (\ref{zbond}) on the PDF of $Z$, we have
\begin{align}
f_V(v)
&    \geq \frac{(R-1)\Gamma(R+1) }{R^R\Gamma(|\mathcal{R}|)} \! \int_1^{vP}\!\!\left(\frac{v}{x} - \frac{1}{P}\right)^{|\mathcal{R}|-1} \!\! e^{-\left(\frac{v}{x} - \frac{1}{P}\right)} \frac{(x-1)^{R-2}}{x^{R+1}} \mathrm{d}x \\
& \label{coff}    =  \frac{R(R-1) }{\Gamma(|\mathcal{R}|)} \int_0^{\infty}\left(\frac{vP-1}{(1+w)P}\right)^{|\mathcal{R}|-1} e^{-\frac{vP-1}{(1+w)P}} \left(\frac{w(vP-1)}{vP(1+w)}\right)^{R+1}\left( \frac{w+vP}{w(vP-1)}\right)^3 \mathrm{d}w \\
&   \label{cofftansonra} \geq \frac{(R-1)\Gamma(R+1) }{R^R\Gamma(|\mathcal{R}|)} \left(\frac{vP-1}{vP}\right)^{R+|\mathcal{R}|-3}e^{-v}v^{|\mathcal{R}|-1}\int_0^{\infty} \frac{w^{R-2}}{(1+w)^{R+|\mathcal{R}|}}\mathrm{d}w,
\end{align}
where we have applied a change of variables $w = \tfrac{vP(x-1)}{vP-x}$ to obtain (\ref{coff}), and (\ref{cofftansonra}) follows from the facts that
$\exp(-\frac{vP-1}{(1+w)P}) = \exp(-v\frac{vP-1}{(1+w)vP}) \geq \exp(-v)$, and $(\frac{w+vP}{w(vP-1)})^3 \geq \frac{1}{w^3}(\frac{vP}{vP-1})^3$.

The integral in (\ref{cofftansonra}) is non-zero and finite for $R\geq 2$. Thus,
\begin{align}
\label{fvlbound}   f_V(v) \geq \letch \Bigl(\frac{vP-1}{vP}\Bigr)^{R+|\mathcal{R}|-3}e^{-v}v^{|\mathcal{R}|-1},
\end{align}
for some constant $0<\letch<\infty$.

Combining (\ref{mainlbound}), (\ref{fvlbound}), and using the fact that $\mathrm{Q}(x) \geq \frac{1}{\sqrt{2\pi}}\frac{x}{1+x^2}\exp(-\frac{x^2}{2})$, we have
\begin{align}
\mathtt{SER}_P(\mathcal{Q}^{\star}_{P, \mathcal{C}}) &    \geq \frac{\letch }{\sqrt{2\pi}} \int_{\frac{1}{P}}^{\infty} \! v^{|\mathcal{R}|-1}e^{-v} \left(\frac{vP-1}{vP}\right)^{R+|\mathcal{R}|-3} \frac{\sqrt{2\letcf vP}}{1+2\letcf vP}e^{-v\letcf P}\mathrm{d}v \\
 & =   \frac{\letch e^{-\letcf -\frac{1}{P}}}{\sqrt{\pi}P^{|\mathcal{R}|}} \int_{0}^{\infty} e^{-w(\letcf +\frac{1}{P})}\frac{ w^{R+|\mathcal{R}|-3}}{(1+w)^{R-2}}\frac{\sqrt{\letcf (1+w)}}{2\letcf w+1+2\letcf } \mathrm{d}w \\
&   \label{buintegraliburadakullan} \geq \frac{\letch \sqrt{\letcf }e^{-1-\letcf }}{(1+2\letcf )\sqrt{\pi}P^{|\mathcal{R}|}} \int_{0}^{\infty} e^{-2\letcf w}w^{R+|\mathcal{R}|-3}(1+w)^{\frac{1}{2}-R} \mathrm{d}w,
\end{align}
where the equality follows from a change of variables $w = vP-1$. The second inequality follows from $\letcf  \geq 1$, and the assumption that $P>1$. Now let us find a lower bound for the integral in (\ref{buintegraliburadakullan}), i.e. $I \triangleq \int_0^{\infty} e^{-\alpha w} w^{\beta} (1+w)^{-\gamma} \mathrm{d}w$, where $\alpha = 2\letcf$, $\beta = R+|\mathcal{R}|-3$, and $\gamma = R - \frac{1}{2}$. Note that $\alpha,\,\gamma > 0$ and $\beta \geq 0$. We have
\begin{align}
  I & =  \int_0^{1} e^{-\alpha w} w^{\beta} (1+w)^{-\gamma} \mathrm{d}w + \int_1^{\infty} e^{-\alpha w} w^{\beta} (1+w)^{-\gamma} \mathrm{d}w \\
&   \geq e^{-\alpha} 2^{-\gamma} \int_0^{1}  w^{\beta} \mathrm{d}w + 2^{-\gamma} \int_1^{\infty} e^{-\alpha w} w^{\beta-\gamma} \mathrm{d}w \\
&   \geq \frac{e^{-\alpha} 2^{-\gamma}}{1+\beta}+ 2^{-\gamma} \int_1^{\infty} e^{-\alpha w} w^{-\beta-\gamma} \mathrm{d}w \\
&   \geq \frac{e^{-\alpha} 2^{-\gamma}}{1+\beta}+ 2^{-\gamma} \int_1^{\infty} e^{-\alpha w} e^{-w(\beta+\gamma)} \mathrm{d}w \\
&   = \frac{e^{-\alpha} 2^{-\gamma}}{1+\beta}+ \frac{2^{-\gamma}  e^{-(\alpha+\gamma+\beta)}}{\alpha + \gamma + \beta}\\
\label{buintegralbudur} & \geq   \frac{2^{1-\gamma}  e^{-(\alpha+\gamma+\beta)}}{1 + \alpha + \gamma + \beta}
\end{align}
Substituting the values of $\alpha,\,\beta$ and $\gamma$ to (\ref{buintegralbudur}), and combining with (\ref{buintegraliburadakullan}), we have
\begin{align}
\mathtt{SER}_P(\mathcal{Q}^{\star}_{P, \mathcal{C}})
&   \geq \frac{\letch \sqrt{\letcf }e^{-1-\letcf }}{(1+2\letcf )\sqrt{\pi}P^{|\mathcal{R}|}} \frac{2^{\frac{3}{2}-R}  e^{-(2\letcf +2R+|\mathcal{R}|-\frac{7}{2})}}{2\letcf +2R+|\mathcal{R}|-\frac{5}{2}}
\end{align}
Since $C_5 \geq 1$, and $2R + |\mathcal{R}| - \frac{5}{2} > 0$ for all $R \geq 2$ and $1 \leq |\mathcal{R}| \leq R$, we have
\begin{align}
\label{yettigayri}
\mathtt{SER}_P(\mathcal{Q}^{\star}_{P, \mathcal{C}}) &   \geq \frac{\letch \sqrt{\letcf }e^{-1-\letcf }}{3\letcf\sqrt{\pi}P^{|\mathcal{R}|}} \frac{2^{\frac{3}{2}-R}  e^{-(2\letcf +2R+R-\frac{7}{2})}}{\letcf(2R+|\mathcal{R}|-\frac{1}{2})}.
\end{align}
Finally, using the fact that $2R+|\mathcal{R}|-\frac{1}{2} \leq 3R$ on the denominator of the second fraction in (\ref{yettigayri}), we can show that (\ref{mainlemmaeq1}) holds for any $R$ with constants $\letca = \min\{\letcpref, \letch \smash\letcg^{-3/2}\,2^{3/2-R}\,\exp(-3R+\frac{5}{2})/(9R\sqrt{\pi})\}$ and $\letcb = 3\letcg$ that are independent of $\mathcal{C}$ and $P$. This concludes the proof.
\section{Proof of Theorem \ref{ldtheorem}}
\label{proofofldtheorem}
We carry out the proof in two parts: First we prove (\ref{rateofconv}), and then the limit arguments in the statement of the theorem.
\subsection{Proof of (\ref{rateofconv})}
For notational convenience, let $\vartheta_P \triangleq  \mathtt{SER}_P(\mathcal{Q}_{P,\mathcal{D}_P}^{\star})$. Also, let $\mathscr{T} = \{\{1,\ldots,R\}-\{r\},r=1,\ldots,R\}$. Using Lemma \ref{mainlemma}, we have
\begin{align}
\label{logic1}
\forall P\geq\letpa,\,\forall\mathcal{T}\in\mathscr{T},\,\biggl\{\xi(\mathcal{D}_P, \mathcal{T}) > \frac{2 C_1}{ \log P}
\implies \vartheta_P >   \frac{\letca (2C_1)^{\frac{3}{2}}}{P^{|\mathcal{T}|+\frac{1}{2}} \log^{\frac{3}{2}}P }\biggr\}.
\end{align}
This, using logical transposition, is equivalent to
\begin{align}
\label{logic2}
\forall P\geq\letpa,\forall \mathcal{T}\in\mathscr{T},\,\biggl\{\vartheta_P \leq   \frac{\letca (2C_1)^{\frac{3}{2}}}{P^{|\mathcal{T}|+\frac{1}{2}} \log^{\frac{3}{2}}P } \implies
\xi(\mathcal{D}_P, \mathcal{T}) \leq \frac{2C_1}{ \log P} \biggr\}.
\end{align}
Since $\vartheta_P \leq   \frac{\letca (2C_1)^{\frac{3}{2}}}{P^{R-\frac{1}{2}} \log^{\frac{3}{2}}P } \implies \vartheta_P \leq   \frac{\letca (2C_1)^{\frac{3}{2}}}{P^{|\mathcal{T}|+\frac{1}{2}} \log^{\frac{3}{2}}P },\,\forall P \geq 1$, it follows from (\ref{logic2}) that
\begin{align}
\label{logic3}
\forall P\geq\letpb,\forall \mathcal{T}\in\mathscr{T},\,\biggl\{\vartheta_P \leq   \frac{\letca (2C_1)^{\frac{3}{2}}}{P^{R-\frac{1}{2}} \log^{\frac{3}{2}}P } \implies
\xi(\mathcal{D}_P, \mathcal{T}) \leq \frac{2C_1}{ \log P} \biggr\},
\end{align}
where $\letpb \triangleq \max\{1,\letpa\}$. It was shown in \cite{koyuncu1} that
\begin{align}
\forall\mathcal{C}_{\mathtt{SRS}} \in\mathscr{C}_{\mathtt{SRS}},\,\mathtt{SER}_P(\mathcal{Q}_{P\!,\,\mathcal{C}_{\mathtt{SRS}}}^{\star}) \leq \letcd P^{-R},\,\forall P \geq \letpc, 
\end{align}
where $0<\letcd,\letpc<\infty$ are constants that are independent of $P$. This upper bound on the SER with $\mathcal{C}_{\mathtt{SRS}}$ holds for any optimal codebook of cardinality at least $R$. Thus, 
\begin{align}
\label{combobmmbob1}
\vartheta_P \leq \letcd P^{-R},\,\forall P \geq \letpc. 
\end{align}
Moreover, there exists a constant $0 < \letpd < \infty$ that is independent of $P$ s.t.
\begin{align}
\label{combobmmbob2}
\letcd P^{-R} \leq  \frac{\letca (2C_1)^{\frac{3}{2}}}{P^{R-\frac{1}{2}} \log^{\frac{3}{2}}P },\,\forall P\geq\letpd. 
\end{align}
Combining (\ref{combobmmbob1}) and (\ref{combobmmbob2}), we have
\begin{align}
\label{combobmmbob3}
\vartheta_P \leq  \frac{\letca (2C_1)^{\frac{3}{2}}}{P^{R-\frac{1}{2}} \log^{\frac{3}{2}}P },\,\forall P\geq\max\{\letpc, \letpd\}.
\end{align}
Letting $\letpe \triangleq \max\{\letpb, \letpc, \letpd\}$, and noting that the left hand side of the implication in (\ref{logic3}) does not depend on $\mathcal{R}$, we have
\begin{align}
\label{logic4}
\forall P\geq\letpe,\,\vartheta_P \leq   \frac{\letca (2C_1)^{\frac{3}{2}}}{P^{R-\frac{1}{2}} \log^{\frac{3}{2}}P } \!\implies\! \forall P\geq\letpe,\,\forall \mathcal{T}\in\mathscr{T},\,
\xi(\mathcal{D}_P, \mathcal{T}) \leq \frac{2 C_1}{\log P}.
\end{align}
According to (\ref{combobmmbob3}), the left hand side of (\ref{logic4}) is true. But (\ref{logic4}) itself is true. We thus have
\begin{align}
\label{logic5}
\forall P\geq\letpf,\,\forall \mathcal{T}\in\mathscr{T},\,
\xi(\mathcal{D}_P, \mathcal{T}) \leq \frac{2C_1}{\log P},
\end{align}
where $\letpf \triangleq \max\{\letpe, \exp(\frac{2C_1}{0.6})\}$. Note that we have also further restricted the power levels that we consider by choosing $P > \exp(\frac{2C_1}{0.6})$ so that $\frac{2C_1}{\log P} < 0.6$.

Now, consider a fixed $\letpg \geq \letpf$. According to (\ref{logic5}),
\begin{align}
\label{logic6}
\forall r\in\{1,\ldots,R\},\,\exists \widetilde{\mathbf{e}}_r\in\mathcal{D}_{\letpg},\,\forall q\in\{1,\ldots,R\}-\{r\},\,|\widetilde{e}_{rq}|^2 \leq \frac{2C_1}{\log P} + \epsilon,
\end{align}
where $\epsilon > 0$ can be arbitrary and $\widetilde{e}_{rq}$ represents the $q$th component of $\widetilde{\mathbf{e}}_r$. Let us choose $\epsilon = \frac{C_1}{\log P}$. Note that, with this choice of $\epsilon$, any $|\widetilde{e}_{rq}|^2$ in (\ref{logic6}) satisfies $|\widetilde{e}_{rq}|^2 \leq 0.9 < 1$.

We now show by contradiction that $\widetilde{\mathbf{e}}_i \neq \widetilde{\mathbf{e}}_j$ whenever $i \neq j$. Suppose that $\widetilde{\mathbf{e}}_i = \widetilde{\mathbf{e}}_j$ with $i\neq j$. Then, $|\widetilde{e}_{ir}|^2 < 1,\,\forall r$, which contradicts the optimality of $\mathcal{D}_{\letpg}$ due to Proposition \ref{codebookoptimalitylemma}. Therefore, for any $\letpg \geq \letpf$, there should be $R$ distinct vectors $\widetilde{\mathbf{e}}_r,\,r=1,\ldots,R$ in $\mathcal{D}_{\letpg}$ with the $r$th satisfying $|\widetilde{e}_{rq}|^2 \leq \frac{3C_1}{\log P},\,\forall q\in\{1,\ldots,R\}-\{r\}$. This concludes the proof of (\ref{rateofconv}).
\subsection{Proof of the Limit Arguments}
We can now prove the limit arguments in the statement of the theorem using (\ref{rateofconv}).

Let $\mathcal{E} = \{\mathbf{x}\in\mathcal{X} : |x_1|=1\}$ represent the set of all SRS vectors that selects the first relay. First, we show that $\exists\mathbf{e}\in\mathcal{E}$ s.t. $\mathbf{e}\in\limsup_{P\rightarrow\infty}\mathcal{D}_P$.

Using (\ref{rateofconv}), we have
\begin{align}
\forall P > \letpththree,\,\exists\mathbf{e}\in\mathcal{E}\mbox{ s.t. }d_P(\mathbf{e}) & = \min_{\mathbf{y}\in\mathcal{D}_P} \|\mathbf{e}-\mathbf{y}\| \leq \|\mathbf{e}-\widetilde{\mathbf{e}}_1\| \leq \sqrt{\frac{3C_1(R-1)}{\log P}},
\end{align}
and therefore, $\min_{\mathbf{e}\in\mathcal{E}} d_P(\mathbf{e}) \leq \sqrt{\frac{3C_1(R-1)}{\log P}}$. It follows that
\begin{align}
\label{limminsifir}
\lim_{P\rightarrow\infty} \min_{\mathbf{e}\in\mathcal{E}} d_P(\mathbf{e}) = 0.
\end{align}

Now, let $\mathbf{e}_P^{\star} \triangleq \min_{\mathbf{e}\in\mathcal{E}} d_P(\mathbf{e})$, and $\mathbf{e}_n^{\star},\,n\in\mathbb{N}$ be a sequence of beamforming vectors. Since $\mathbf{e}_n^{\star}\in\mathcal{E},\,\forall n$ and $\mathcal{E}$ is compact, by the Bolzano-Weierstrass theorem, the sequence $\mathbf{e}_n^{\star},\,n\in\mathbb{N}$ contains a subsequence $\mathbf{e}_{n_i}^{\star},\,i\in\mathbb{N}$ with $\lim_{i\rightarrow\infty} \mathbf{e}_{n_i}^{\star} = \mathbf{e}^{\star}$ for some $\mathbf{e}^{\star}\in\mathcal{E}$.

Note that for any $\mathbf{y}\in\mathbb{C}^R$, $
\|\mathbf{y} - \mathbf{e}^{\star}\| \leq \|\mathbf{y} - \mathbf{e}_{n_i}\| + \|\mathbf{e}^{\star} - \mathbf{e}_{n_i}\|$ by triangle inequality. It follows that
$\min_{\mathbf{y}\in\mathcal{D}_{n_i}}\|\mathbf{y} - \mathbf{e}^{\star}\| \leq \min_{\mathbf{y}\in\mathcal{D}_{n_i}}\left\{\|\mathbf{y} - \mathbf{e}_{n_i}\| + \|\mathbf{e}^{\star} - \mathbf{e}_{n_i}\|\right\} = \min_{\mathbf{y}\in\mathcal{D}_{n_i}}\|\mathbf{y} - \mathbf{e}_{n_i}\| + \|\mathbf{e}^{\star} - \mathbf{e}_{n_i}\|$. Rearranging the terms, we have
$d_{n_i}(\mathbf{e}^{\star}) - d_{n_i}(\mathbf{e}_{n_i}^{\star}) \leq \| \mathbf{e}^{\star} - \mathbf{e}_{n_i}^{\star}\|,\,\forall i\in\mathbb{N}$, and thus
\begin{align}
\liminf_{i\rightarrow\infty} \left(d_{n_i}(\mathbf{e}^{\star}) - d_{n_i}(\mathbf{e}_{n_i}^{\star}) \right) & \leq \liminf_{i\rightarrow\infty}\|\mathbf{e}^{\star} - \mathbf{e}_{n_i}^{\star}\| \\ \label{budabaskbirliminf} & = 0.
\end{align}
The equality follows from the fact that $\lim_{i\rightarrow\infty} \mathbf{e}_{n_i}^{\star} = \mathbf{e}^{\star}$.

We now have
\begin{align}
\liminf_{P\rightarrow\infty} d_{P}(\mathbf{e}^{\star}) - \limsup_{P\rightarrow\infty} d_{P}(\mathbf{e}_{P}^{\star}) & = \liminf_{P\rightarrow\infty} d_{P}(\mathbf{e}^{\star}) + \liminf_{P\rightarrow\infty} \left(-d_{P}(\mathbf{e}_{P}^{\star})\right) \\
& \label{liminffirsteq} \leq \liminf_{P\rightarrow\infty}\left(d_{P}(\mathbf{e}^{\star}) - d_{P}(\mathbf{e}_{P}^{\star})  \right) \\
& \label{liminfsecondeq} \leq \liminf_{i\rightarrow\infty} \left(d_{n_i}(\mathbf{e}^{\star}) - d_{n_i}(\mathbf{e}_{n_i}^{\star})\right), \\
\label{liminfinfinalinequalitisbu} & \leq 0,
\end{align}
where (\ref{liminffirsteq}) follows since 
\begin{align}
\label{thatliminfbound}
\liminf_{x\rightarrow\infty}f(x)+\liminf_{x\rightarrow\infty}g(x) \leq \liminf_{x\rightarrow\infty}\left(f(x) + g(x)\right),
\end{align}
for any functions $f$ and $g$. For (\ref{liminfsecondeq}), we have used the fact that the lower limit of a sequence is less than the lower limit of any of its subsequences. For (\ref{liminfinfinalinequalitisbu}), we have used (\ref{budabaskbirliminf}).

Now, since $\liminf_{P\rightarrow\infty} d_{P}(\mathbf{e}^{\star}) - \limsup_{P\rightarrow\infty} d_{P}(\mathbf{e}_{P}^{\star}) \leq 0$ as shown in the derivation above, we have
\begin{align}
\liminf_{P\rightarrow\infty} d_{P}(\mathbf{e}^{\star}) \leq \limsup_{P\rightarrow\infty} d_{P}(\mathbf{e}_{P}^{\star}) = \limsup_{P\rightarrow\infty} \min_{\mathbf{e}\in\mathcal{E}} d_P(\mathbf{e})  = 0,
\end{align}
where the last equality follows from (\ref{limminsifir}). Therefore, $\liminf_{P\rightarrow\infty} d_{P}(\mathbf{e}^{\star})  \leq 0$. On the other hand, obviously we have $\liminf_{P\rightarrow\infty} d_{P}(\mathbf{e}^{\star}) \geq 0$. Combining the two inequalities yields $\liminf_{P\rightarrow\infty} d_{P}(\mathbf{e}^{\star}) = 0$. This shows the existence of an SRS vector $\mathbf{e}\in\mathcal{E}$ (namely $\mathbf{e}^{\star}$) that selects the first relay and satisfies $\liminf_{P\rightarrow\infty} d_P(\mathbf{e}) = 0$, or equivalently, $\mathbf{e}\in\limsup_{P\rightarrow\infty}\mathcal{D}_P$. We can similarly show the existence of the remaining $R-1$ SRS vectors in the upper limit of $\mathcal{D}$. Therefore, $\exists\mathcal{C}_{\mathtt{SRS}}\in\mathscr{C}_{\mathtt{SRS}}$ s.t. $\mathcal{C}_{\mathtt{SRS}}\subset\limsup_{P\rightarrow\infty}\mathcal{D}_P$.

What is left is to show that if $|\mathcal{D}| = R$, and $\lim_{P\rightarrow\infty}\mathcal{D}_P$ exists, we have $\exists\mathcal{C}_{\mathtt{SRS}}'\in\mathscr{C}_{\mathtt{SRS}}$ s.t. $\mathcal{C}_{\mathtt{SRS}}'=\lim_{P\rightarrow\infty}\mathcal{D}_P$. We have shown that $\mathcal{C}_{\mathtt{SRS}}' \subset \limsup_{P\rightarrow\infty}\mathcal{D}_P$. If $\lim_{P\rightarrow\infty}\mathcal{D}_P$ exists, then $\limsup_{P\rightarrow\infty}\mathcal{D}_P = \lim_{P\rightarrow\infty}\mathcal{D}_P$, and thus $\mathcal{C}_{\mathtt{SRS}}' \subset \lim_{P\rightarrow\infty}\mathcal{D}_P$ with $|\lim_{P\rightarrow\infty}\mathcal{D}_P| \geq R$. To complete the proof, it is therefore sufficient to show that $|\lim_{P\rightarrow\infty}\mathcal{D}_P| \leq R$.

The following lemma shows that when $\lim_{P\rightarrow\infty}\mathcal{D}_P$ exists, its cardinality cannot be more than $|\mathcal{D}|$, and thus concludes the proof of the theorem.
\begin{lemma}
\label{limitexistssebokolur}
For any d-codebook $\mathcal{D}$ with $|\mathcal{D}| < \infty$, if $\lim_{P\rightarrow\infty}\mathcal{D}_P$ exists, then $|\lim_{P\rightarrow\infty}\mathcal{D}_P| \leq |\mathcal{D}|$.
\end{lemma}
\begin{proof}
Let $\mathcal{L} = \lim_{P\rightarrow\infty}\mathcal{D}_P$. Suppose that $|\mathcal{L}| \geq |\mathcal{D}|+1$. Then, $\exists\mathbf{x}_i,\ldots,\mathbf{x}_{|\mathcal{D}|+1}\in\mathcal{L}$, with $\forall i,j\in\{1,\ldots,|\mathcal{D}|+1\},\,\mathbf{x}_i \neq \mathbf{x}_j \Longleftrightarrow i \neq j$. Since $\mathcal{L} = \liminf_{P\rightarrow\infty}\mathcal{D}_P$ as well, we have
$\forall i\in\{1,\ldots, |\mathcal{D}| + 1\}$, $\lim_{P\rightarrow\infty}\min_{\mathbf{y}\in\mathcal{D}_P} \|\mathbf{y} - \mathbf{x}_i\| = 0$ by the definition of $\liminf_{P\rightarrow\infty}\mathcal{D}_P$. This implies that $\forall i\in\{1,\ldots,|\mathcal{D}| + 1\},\,\forall \epsilon > 0,\,\exists P_{i,\epsilon} > 0\mbox{ s.t. }\forall P > P_{i,\epsilon},\,\min_{\mathbf{y}\in\mathcal{D}_P}\|\mathbf{y} - \mathbf{x}_i\| \leq \epsilon$. Letting $P_{\epsilon} = \max_i P_{i,\epsilon}$, we have
\begin{align}
\label{limitargument}
\forall \epsilon > 0,\,\exists P_{\epsilon} > 0\mbox{ s.t. }\forall P > P_{\epsilon},\,\forall i\in\{1,\ldots,|\mathcal{D}| + 1\},\,\min_{\mathbf{y}\in\mathcal{D}_P}\|\mathbf{y} - \mathbf{x}_i\| \leq \epsilon. \end{align}

Now, let
\begin{align}
\label{deltaargument}
\delta = \min_{\substack{i,j\in\{1,\ldots,|\mathcal{D}|+1\}\\i \neq j}} \|\mathbf{x}_i - \mathbf{x}_j\|,
\end{align}
set $\epsilon = \delta/4$, and consider a fixed $P_0 > P_{\delta/4}$. Also, let $\mathbf{y}_i = \arg\min_{\mathbf{y}\in\mathcal{D}_{P_0}} \|\mathbf{y} - \mathbf{x}_i\|,\,i=1,\ldots,|\mathcal{D}|+1$. Since $|\mathcal{D}_{P_0}| = |\mathcal{D}|$, $\mathbf{y}_k = \mathbf{y}_\ell \triangleq \widetilde{\mathbf{y}}$ for some $k \neq \ell$. Note that, as a result of (\ref{limitargument}) and the definition of $\widetilde{\mathbf{y}}$, we have $\|\widetilde{\mathbf{y}} - \mathbf{x}_k\| \leq \delta/4$ and $\|\widetilde{\mathbf{y}} - \mathbf{x}_\ell\| \leq \delta/4$.

However, by triangle inequality, $\|\mathbf{x}_k - \mathbf{x}_\ell\| \leq \|\mathbf{x}_k - \widetilde{\mathbf{y}}\| + \|\mathbf{x}_\ell - \widetilde{\mathbf{y}}\| \leq \delta/4+\delta/4 = \delta/2$, and this contradicts (\ref{deltaargument}). Therefore, the cardinality of $\mathcal{L}$ cannot be more than $|\mathcal{D}|$, concluding the proof.
\end{proof}
\section{Proof of Theorem \ref{sdtheorem}}
\label{proofofsdtheorem}
Similar to what has been done in the proof of Theorem \ref{ldtheorem}, we carry out the proof in two parts: First we prove (\ref{sdtheoremeq}), and then the limit arguments in the statement of the theorem.
\subsection{Proof of (\ref{sdtheoremeq})}
Let $\mathscr{T}$ represent the collection of all subsets of $\{1,\ldots,R\}$ with cardinality no greater than $|\mathcal{D}|-1$. Then, using the same ideas as in the proof of Theorem \ref{ldtheorem}, we have
\begin{align}
\label{condgaribus}
\forall P>\sdtheorempa,\,\forall \mathcal{T}\in\mathscr{T},\,
\xi(\mathcal{D}_P, \mathcal{T}) \leq \frac{\sdtheoremca}{\log P},
\end{align}
for constants $0 < \sdtheoremca,\, \sdtheorempa <\infty$ independent of $P$. Let us set the constants in the statement of the theorem as $\sdtheoremcb = R\sdtheoremca$ and $\sdtheorempb = \sdtheorempa$.
We now prove (\ref{sdtheoremeq}) by contradiction. Suppose that (\ref{sdtheoremeq}) is false. Then, $\exists \sdtheorempc > \sdtheorempb = \sdtheorempa,\,\exists\mathbf{x},\mathbf{y}\in\mathcal{C},\,\mathbf{x}\neq\mathbf{y},\,\sum_{r=1}^R |x_r||y_r| > \frac{ R\sdtheoremca}{\log P}$. This implies that $\exists r\in\{1,\ldots,R\},\,|x_r||y_r| > \frac{ \sdtheoremca}{\log P}$. Hence, either we have $|x_r|> (\frac{\sdtheoremca}{\log P})^{1/2}$ or $|y_r| > (\frac{\sdtheoremca}{\log P})^{1/2}$.

Now, let $\mtf = \{r\}\cup\{\iota(\mathbf{z}): \mathbf{z}\in\mathcal{D}_{\sdtheorempc} - \{\mathbf{x},\mathbf{y}\}\}$, where $\iota(\mathbf{z})$ is any index that satisfies $|z_{\iota(\mathbf{z})}| = 1$. Since $|\mtf| \leq |\mathcal{D}| - 1$, $\mtf\in\mathscr{T}$. Also, either $\max_{t\in\mtf}|x_t|^2> \frac{\sdtheoremca}{\log P}$ or $\max_{t\in\mtf}|y_t|^2 > \frac{\sdtheoremca}{\log P}$. Moreover, $\forall \mathbf{z}\in\mathcal{D}_{\sdtheorempc} - \{\mathbf{x},\mathbf{y}\},\,\max_{t\in\mtf} |z_t|^2 = 1$ by the construction of $\mtf$. Therefore, $\xi(\mathcal{D}_{\sdtheorempc}, \mtf) = \inf_{\mathbf{x}\in\mathcal{D}_{\sdtheorempc}} \max_{t\in\mtf} |x_t|^2 > \frac{\sdtheoremca}{\log P}$. But, this contradicts (\ref{condgaribus}), and thus concludes the proof of (\ref{sdtheoremeq}).
\subsection{Proof of the Limit Arguments}
We can now prove the limit arguments in the statement of the theorem using (\ref{sdtheoremeq}).

Let $\mathcal{O} = \{\mathbf{o}_1, \ldots,\mathbf{o}_{|\mathcal{D}|}\}$ represent an OMRS codebook given a beamforming codebook cardinality $1 \leq |\mathcal{D}| \leq R$. For any OMRS codebook $\mathcal{O}$, we define its ``vectorized'' version $\mathtt{vec}(\mathcal{O}) \triangleq [\mathbf{o}_1\cdots\mathbf{o}_{|\mathcal{D}|}]$ as an alternative representation for $\mathcal{O}$. Also, let $\mathfrak{O}\triangleq\bigcup_{\mathcal{O}\in\mathscr{O}}\mathtt{vec}(\mathcal{O})$ represent the collection of all vectorized OMRS codebooks.

We now need the following lemma to proceed:
\begin{lemma}
For any $\mathcal{D}_P$ that satisfies (\ref{sdtheoremeq}), $\exists\mathbf{o}\in\mathfrak{O}$ s.t. $\min_{\mathbf{y}\in\mathcal{D}_P^{|\mathcal{D}|}} \|\mathbf{o} - \mathbf{y}\| \leq \frac{\sdtheoremcb\sqrt{R|\mathcal{D}|}}{\log P}$.
\end{lemma}
\begin{proof}
Let $\mathcal{D}_P = \{\mathbf{x}_1, \ldots, \mathbf{x}_{|\mathcal{D}|}\}$. For convenience, we rewrite the condition in (\ref{sdtheoremeq}) as
\begin{align}
\label{orthocond}
\max_{\substack{i,j\in\{1,\ldots,|\mathcal{D}|\} \\ i \neq j}} \sum_{r=1}^R |x_{ir}||x_{jr}| \leq \delta,
\end{align}
where $\delta = \frac{\sdtheoremcb}{\log P}$. Now, (\ref{orthocond}) implies that for all $i,j\in\{1,\ldots,|\mathcal{D}|\}$ with $i \neq j$  we have $|x_{ir}||x_{jr}| \leq \delta,\,\forall r\in\{1,\ldots,R\}$. Then, given any $r\in\{1,\ldots,R\}$, either $|x_{ir}| \leq \sqrt{\delta},\,\forall i\in\{1,\ldots,|\mathcal{D}|\}$, or there exists only one index $i_r^{\star}\in\{1,\ldots,|\mathcal{D}|\}$ s.t. $|x_{i_r^{\star}, r}|> \sqrt{\delta}$, and $|x_{ir}| \leq \sqrt{\delta},\,\forall i\in\{1,\ldots,|\mathcal{D}|\}-\{i_r^{\star}\}$.

Let us now define the function $f:\mathbb{C}\rightarrow\mathbb{C}$ with $f(x) = x$ if $|x| > \sqrt{\delta}$, and $f(x) = 0$, otherwise. Now, let $o_{ir} = f(x_{ir}),\,i=1,\ldots,|\mathcal{D}|,\,r=1,\ldots,R$. As a result of the properties of $f$ and $x_{ir}$, not only $\mathbf{o}\in\mathfrak{O}$ is an OMRS codebook, but also,
\begin{align}
\min_{\mathbf{y}\in\mathcal{D}_P^{|\mathcal{D}|}} \|\mathbf{o} - \mathbf{y}\| & = \min_{\mathbf{y}\in\mathcal{D}_P^{|\mathcal{D}|}} \sqrt{\sum_{r=1}^R \sum_{i=1}^{|\mathcal{D}|} |o_{ir} - y_{ir}|^2} \\ & \leq \sqrt{\sum_{r=1}^R\sum_{i=1}^{|\mathcal{D}|} |o_{ir} - x_{ir}|^2} \\ & \leq \sqrt{\sum_{r=1}^R |\mathcal{D}|\delta^2 } \\ & = \sqrt{R|\mathcal{D}|} \delta,
\end{align}
and this concludes the proof.
\end{proof}
In other words, if $\mathcal{D}$ is an optimal d-codebook, for all $P$ sufficiently large, we can find an OMRS codebook $\mathbf{o}\in\mathfrak{O}$ s.t. $\mathbf{o}$ is as close as $\frac{\sdtheoremcb\sqrt{R|\mathcal{D}|}}{\log P}$ to $\mathcal{D}_P$. We thus have
\begin{align}
\label{infimumluyaklasim}
\lim_{P\rightarrow\infty} \inf_{\mathbf{o}\in\mathcal{O}} \min_{\mathbf{y}\in\mathcal{D}_P^{|\mathcal{D}|}} \|\mathbf{y} - \mathbf{o}\|= 0.
\end{align}

We now prove that $\mathfrak{O}$ is a compact set so that we can replace the infimum in (\ref{infimumluyaklasim}) by a minimum.
\begin{lemma}
$\mathfrak{O}$ is compact.
\end{lemma}
\begin{proof}
It is sufficient to show that $\mathfrak{O}$ is bounded and closed. Since $\mathfrak{O} \subset \mathcal{X}^{|\mathcal{D}|}$ and $\mathcal{X}^{|\mathcal{D}|}$ is bounded,  $\mathfrak{O}$ is bounded. We prove that $\mathfrak{O}$ is also closed by showing that it can be expressed as the union of a finite number of closed sets. First, we need the following definitions:

\begin{itemize}
\item Let $\mathcal{V}$ represent the set of all vectors $[\begin{array}{cccccc} \alpha_1 & \cdots & \alpha_{|\mathcal{D}|} & \beta_1 & \cdots & \beta_{R - |\mathcal{D}|}\end{array}]$ that satisfy the following:
\begin{enumerate}
\item $\alpha_1,\,\ldots,\,\alpha_{|\mathcal{D}|},\,\beta_1,\,\ldots,\,\beta_{R - |\mathcal{D}|}$ are positive integers.
\item $1 \leq \alpha_1,\ldots,\alpha_{|\mathcal{D}|} \leq R$.
\item $\forall i,j\in\{1,\ldots,|\mathcal{D}|\},\,\alpha_i \neq \alpha_j \Leftrightarrow i\neq j$.
\item $1 \leq \beta_1, \ldots, \beta_{R - |\mathcal{D}|} \leq |\mathcal{D}|$.
\end{enumerate}
Note that $|\mathcal{V}| = R(R-1)\cdots(R-|\mathcal{D}|+1)|\mathcal{D}|^{R - |\mathcal{D}|}$.

\item Let $\mathtt{DISK} = \{x\in\mathbb{C}:\|x\|\leq 1\}$ and $\mathtt{CIRC} = \{x\in\mathbb{C}:\|x\|=1\}$ represent the unit disk and the unit circle, respectively.
\item Given $\mathbf{v} = [\begin{array}{cccccc} \alpha_1 & \cdots & \alpha_{|\mathcal{D}|} & \beta_1 & \cdots & \beta_{R - |\mathcal{D}|}\end{array}]\in\mathcal{V}$, let $\mathfrak{X}_{\mathbf{v}}$ represent the collection of all vectorized codebooks $[\begin{array}{ccc}\mathbf{x}_1 & \cdots & \mathbf{x}_{|\mathcal{D}|}\end{array}] = [\begin{array}{ccccccc} x_{11} & \cdots & x_{1R} & \cdots & x_{|\mathcal{D}|,1} & \cdots & x_{|\mathcal{D}|, R}\end{array}]$ with the following properties:
\begin{enumerate}
\item $\forall i\in\{1,\ldots,|\mathcal{D}|\},\,x_{i\alpha_i} \in \mathtt{CIRC},\,\forall j\in\{1,\ldots,|\mathcal{D}|\}-\{i\},\,x_{j\alpha_i} = 0$.
\item $\forall i\in\{1,\ldots,R - |\mathcal{D}|\},\,x_{\beta_i\gamma_i} \in \mathtt{DISC},\,\forall j\in\{1,\ldots,R - |\mathcal{D}|\}- \{\beta_i\},\,x_{j\gamma_i} = 0$, where $\gamma_1 < \cdots < \gamma_{R - |\mathcal{D}|}$ satisfy $\{\gamma_1, \ldots, \gamma_{R - |\mathcal{D}|}\} = \{1,\ldots,R\}-\{\alpha_1, \ldots, \alpha_{|\mathcal{D}|}\}$.
\end{enumerate}
\end{itemize}

According to these properties, for any given $\mathbf{v}\in\mathcal{V}$, $\mathfrak{X}_{\mathbf{v}}$ can be expressed as a finite cartesian product of the closed sets $\mathtt{DISC}$, $\mathtt{CIRC}$ and $\{0\}$. Hence, $\mathfrak{X}_{\mathbf{v}}$ is closed for any $\mathbf{v}$.

It is straightforward to show that $\mathfrak{X}_{\mathbf{v}}$ is a set of OMRS codebooks for any given $\mathbf{v}\in\mathcal{V}$, and thus
$\mathbf{o}\in\bigcup_{\mathbf{v}\in\mathcal{V}} \mathfrak{X}_{\mathbf{v}} \implies \mathbf{o}\in \mathfrak{O}$. Each $\mathbf{v}\in\mathcal{V}$ actually corresponds to a particular OMRS structure. As an example, for $R=2$ and $|\mathcal{D}| = 1$, let $\mathbf{w} = [\begin{array}{cccc} 1 & 3 & 1 & 1 \end{array}]$. Then, $\mathfrak{X}_{\mathbf{w}}$ is the union of all OMRS codebooks of structure $[\begin{array}{cccccccc} x_{11} & x_{12} & 0 & x_{14} & 0 & 0 & x_{23} & 0 \end{array}]$, where $x_{11},x_{23}\in\mathtt{CIRC}$ and $x_{12},x_{14}\in\mathtt{DISC}$.

We now show the converse, i.e. $\mathbf{o}\in \mathfrak{O} \implies \mathbf{o}\in\bigcup_{\mathbf{v}\in\mathcal{V}} \mathfrak{X}_{\mathbf{v}}$. Consider some $\mathbf{o}\in\mathfrak{O}$. We shall construct a $\mathbf{v} = [\begin{array}{cccccc} \alpha_1 & \cdots & \alpha_{|\mathcal{D}|} & \beta_1 & \cdots & \beta_{R - |\mathcal{D}|}\end{array}]\in\mathcal{V}$ s.t. $\mathbf{o}\in\mathfrak{X}_{\mathbf{v}}$. Since $\mathbf{o}$ is an optimal codebook, by Proposition \ref{codebookoptimalitylemma}, at least one component of every beamforming vector in $\mathbf{o}$ has unit norm, and thus we choose the $\alpha_i$ in such a way that $|o_{i\alpha_i}| = 1$, or equivalently, $o_{i\alpha_i}\in\mathtt{CIRC}$. Also, since $\mathbf{o}$ is an OMRS codebook, we have $o_{j\alpha_i} = 0,\,\forall j \neq i$ by definition. This satisfies the first property in the definition of $\mathfrak{X}_{\mathbf{v}}$.

Now, let $\gamma_1 < \cdots < \gamma_{R - |\mathcal{D}|}$ satisfy $\{\gamma_1, \ldots, \gamma_{R - |\mathcal{D}|}\} = \{1,\ldots,R\}-\{\alpha_1, \ldots, \alpha_{|\mathcal{D}|}\}$. For any given $\gamma_i$, there are two possibilities:
\begin{enumerate}
\item $x_{j\gamma_i}=0,\,\forall j\in\{1,\ldots,|\mathcal{D}|\}$. In this case we can pick any $1 \leq \beta_i \leq |\mathcal{D}|$.
\item Since $\mathbf{o}$ is an OMRS codebook, there is at most one non-zero $x_{j\gamma_i},\,j=1,\ldots,|\mathcal{D}|$. Suppose that $x_{j'\gamma_i}\neq 0$. Then, we set $\beta_i = j'$.
\end{enumerate}
This satisfies the second property in the definition of $\mathfrak{X}_{\mathbf{v}}$. Therefore, for the particular $\alpha$s and $\beta$s we have chosen, $\mathbf{o}\in\mathfrak{X}_v$, and thus in general $\mathbf{o}\in \mathfrak{O} \implies \mathbf{o}\in\bigcup_{\mathbf{v}\in\mathcal{V}} \mathfrak{X}_{\mathbf{v}}$. Combining this with the fact that $\mathbf{o}\in\bigcup_{\mathbf{v}\in\mathcal{V}} \mathfrak{X}_{\mathbf{v}} \implies \mathbf{o}\in \mathfrak{O}$, we have $\mathfrak{O} = \bigcup_{\mathbf{v}\in\mathcal{V}} \mathfrak{X}_{\mathbf{v}}$. Hence $\mathfrak{O}$ is the union of a finite number of closed sets. Therefore, it is closed, and this concludes the proof.
\end{proof}
Hence, we can rewrite (\ref{infimumluyaklasim}) as
\begin{align}
\label{minimumluyaklasim}
\lim_{P\rightarrow\infty} \min_{\mathbf{o}\in\mathcal{O}} \min_{\mathbf{y}\in\mathcal{D}_P^{|\mathcal{D}|}} \|\mathbf{y} - \mathbf{o}\|= 0.
\end{align}
Note that this equality has the same form as (\ref{limminsifir}). Using the exact same steps as in Appendix \ref{proofofldtheorem}, we can show that
\begin{align}
\label{minimumluyaklasimmmm}
 \exists\mathbf{o}^{\star}\in\mathfrak{O}\mbox{ s.t. } \liminf_{P\rightarrow\infty} \min_{\mathbf{y}\in\mathcal{D}_P^{|\mathcal{D}|}} \|\mathbf{y} - \mathbf{o}^{\star}\|= 0.
\end{align}

Now, we have
\begin{align}
0 & = \liminf_{P\rightarrow\infty} \min_{\mathbf{y}\in\mathcal{D}_P^{|\mathcal{D}|}} \|\mathbf{y} - \mathbf{o}^{\star}\|\\
\label{hhhhhooooollllddddeeerrrr}& \geq \liminf_{P\rightarrow\infty} \min_{\mathbf{y}\in\mathcal{D}_P^{|\mathcal{D}|}} \frac{1}{\sqrt{|\mathcal{D}|}}\sum_{i=1}^{|\mathcal{D}|}\|\mathbf{y}_i - \mathbf{o}_i^{\star}\| \\
& = \frac{1}{\sqrt{|\mathcal{D}|}} \liminf_{P\rightarrow\infty} \min_{\mathbf{y}_1\in\mathcal{D}_P} \cdots \min_{\mathbf{y}_{|\mathcal{D}|}\in\mathcal{D}_P} \sum_{i=1}^{|\mathcal{D}|}\|\mathbf{y}_i - \mathbf{o}_i^{\star}\| \\
& = \frac{1}{\sqrt{|\mathcal{D}|}} \liminf_{P\rightarrow\infty}  \sum_{i=1}^{|\mathcal{D}|} \min_{\mathbf{y}_i\in\mathcal{D}_P} \|\mathbf{y}_i - \mathbf{o}_i^{\star}\| \\
\label{againliminfineq} & \geq \frac{1}{\sqrt{|\mathcal{D}|}} \sum_{i=1}^{|\mathcal{D}|} \liminf_{P\rightarrow\infty}  \min_{\mathbf{y}_i\in\mathcal{D}_P} \|\mathbf{y}_i - \mathbf{o}_i^{\star}\|,
\end{align}
where (\ref{hhhhhooooollllddddeeerrrr}) follows from H\"{o}lder's inequality. For (\ref{againliminfineq}), we have used (\ref{thatliminfbound}).

Finally, using (\ref{againliminfineq}), $\forall i\in\{1,\ldots,|\mathcal{D}|\}$, we have $\liminf_{P\rightarrow\infty}  \min_{\mathbf{y}\in\mathcal{D}_P} \|\mathbf{y} - \mathbf{o}_i^{\star}\| = 0$, or equivalently $\mathbf{o}_i^{\star} \in \limsup_{P\rightarrow\infty} \mathcal{D}_P$. This shows the existence of an OMRS codebook $\mathcal{O}^{\star}\in\mathscr{O}(|\mathcal{D}|)$ s.t. $\mathcal{O}^{\star}\subset\limsup_{P\rightarrow\infty} \mathcal{D}_P$. If the limit exists, according to Lemma \ref{limitexistssebokolur}, $\mathcal{O}^{\star}=\lim_{P\rightarrow\infty} \mathcal{D}_P$, concluding the proof.
\bibliographystyle{IEEEtran}
\bibliography{IEEEabrv,letter}

\begin{thebibliography}{10}
\providecommand{\url}[1]{#1}
\csname url@samestyle\endcsname
\providecommand{\newblock}{\relax}
\providecommand{\bibinfo}[2]{#2}
\providecommand{\BIBentrySTDinterwordspacing}{\spaceskip=0pt\relax}
\providecommand{\BIBentryALTinterwordstretchfactor}{4}
\providecommand{\BIBentryALTinterwordspacing}{\spaceskip=\fontdimen2\font plus
\BIBentryALTinterwordstretchfactor\fontdimen3\font minus
  \fontdimen4\font\relax}
\providecommand{\BIBforeignlanguage}[2]{{%
\expandafter\ifx\csname l@#1\endcsname\relax
\typeout{** WARNING: IEEEtran.bst: No hyphenation pattern has been}%
\typeout{** loaded for the language `#1'. Using the pattern for}%
\typeout{** the default language instead.}%
\else
\language=\csname l@#1\endcsname
\fi
#2}}
\providecommand{\BIBdecl}{\relax}
\BIBdecl

\bibitem{larsson1}
E.~G. Larsson and Y.~Cao, ``Collaborative transmit diversity with adaptive
  radio resource and power allocation,'' \emph{{IEEE} Commun. Lett.}, vol.~9,
  no.~6, pp. 511--513, Jun. 2006.

\bibitem{jing1}
Y.~Jing and H.~Jafarkhani, ``Network beamforming using relays with perfect
  channel information,'' \emph{{IEEE} Trans. Inf. Theory}, vol.~55, no.~6, pp.
  2499--2517, Jun. 2009.

\bibitem{koyuncu1}
E.~Koyuncu, Y.~Jing, and H.~Jafarkhani, ``Distributed beamforming in wireless
  relay networks with quantized feedback,'' \emph{IEEE J. Select. Areas
  Commun.}, vol.~26, no.~8, pp. 1429--1439, Oct. 2008.

\bibitem{laneman2}
J.~N. Laneman and G.~W. Wornell, ``Distributed space-time-coded protocols for
  exploiting cooperative diversity in wireless networks,'' \emph{{IEEE} Trans.
  Inf. Theory}, vol.~49, no.~10, pp. 2415--2425, Oct. 2003.

\bibitem{jing4}
Y.~Jing and B.~Hassibi, ``Distributed space-time coding in wireless relay
  networks,'' \emph{{IEEE} Trans. Wireless Commun.}, vol.~5, no.~12, pp.
  3524--3536, Dec. 2006.

\bibitem{zhao5}
Y.~Zhao, R.~Adve, and T.~J. Lim, ``Beamforming with limited feedback in
  amplify-and-forward cooperative networks,'' \emph{{IEEE} Trans. Wireless
  Commun.}, vol.~7, no.~12, pp. 5145--5149, Dec. 2008.

\bibitem{bletsas1}
A.~Bletsas, A.~Khisti, D.~P. Reed, and A.~Lippman, ``A simple cooperative
  diversity method based on network path selection,'' \emph{IEEE J. Select.
  Areas Commun.}, vol.~24, no.~3, pp. 659--672, Mar. 2006.

\bibitem{riberio1}
A.~Riberio, X.~Cai, and G.~B. Giannakis, ``Symbol error probabilities for
  general cooperative links,'' \emph{{IEEE} Trans. Wireless Commun.}, vol.~4,
  no.~3, pp. 1264--1273, May 2005.

\bibitem{anghel1}
P.~A. Anghel and M.~Kaveh, ``Exact symbol error probability of a cooperative
  network in a {R}ayleigh-fading environment,'' \emph{{IEEE} Trans. Wireless
  Commun.}, vol.~3, no.~5, pp. 1416--1421, Sep. 2004.

\bibitem{hasna1}
M.~O. Hasna and M.-S. Aoluini, ``Optimal power allocation for relayed
  transmissions over {R}ayleigh-fading channels,'' \emph{{IEEE} Trans. Wireless
  Commun.}, vol.~3, no.~6, pp. 1999--2004, Nov. 2004.

\bibitem{zhao3}
Y.~Zhao, R.~S. Adve, and T.~J. Lim, ``Improving amplify-and-forward relay
  networks: Optimal power allocation versus selection,'' \emph{{IEEE} Trans.
  Wireless Commun.}, vol.~6, no.~8, pp. 3114--3123, Aug. 2007.

\bibitem{zhao6}
Y.~Zhao, R.~Adve, and T.~J. Lim, ``Symbol error rate of selection
  amplify-and-forward relay systems,'' \emph{{IEEE} Commun. Lett.}, vol.~10,
  no.~11, pp. 757--759, Nov. 2006.

\bibitem{jing6}
Y.~Jing and H.~Jafarkhani, ``Single and multiple relay selection schemes and
  their achievable diversity orders,'' \emph{{IEEE} Trans. Wireless Commun.},
  vol.~8, no.~3, pp. 1414--1423, Mar. 2009.

\bibitem{shi1}
W.~Shi and S.~Roy, ``Achieving full diversity by selection in arbitrary
  multi-hop amplify-and-forward relay networks,'' in \emph{IEEE Global
  Telecommun. Conf.}, Dec. 2010.

\bibitem{koyuncu3}
\BIBentryALTinterwordspacing
E.~Koyuncu and H.~Jafarkhani, ``Distributed beamforming in wireless multiuser
  relay-interference networks with quantized feedback,'' \emph{submitted to
  IEEE Trans. Inf. Theory}. [Online]. Available:
  \url{http://arxiv.org/abs/1007.5514}
\BIBentrySTDinterwordspacing

\bibitem{mukkavilli1}
K.~K. Mukkavilli, A.~Sabharwal, and E.~Erkip, ``On beamforming with finite rate
  feedback in multiple-antenna systems,'' \emph{{IEEE} Trans. Inf. Theory},
  vol.~49, no.~10, pp. 2562--2579, Oct. 2003.

\bibitem{roh1}
J.~C. Roh and B.~D. Rao, ``Transmit beamforming in multiple-antenna systems
  with finite rate feedback: A {VQ}-based approach,'' \emph{{IEEE} Trans. Inf.
  Theory}, vol.~52, no.~3, pp. 1101--1112, Mar. 2006.

\bibitem{love1}
D.~J. Love and {R. W. Heath, Jr.}, ``Grassmannian beamforming for
  multiple-input multiple-output wireless systems,'' \emph{{IEEE} Trans. Inf.
  Theory}, vol.~49, no.~10, pp. 2735--2747, Oct. 2003.

\bibitem{koyuncu2}
E.~Koyuncu and H.~Jafarkhani, ``A systematic distributed quantizer design
  method with an application to {MIMO} broadcast channels,'' in \emph{{IEEE}
  Data Commun. Conf.}, Mar. 2010.

\bibitem{sva}
J.-P. Aubin and H.~Frankowska, \emph{Set-valued analysis}.\hskip 1em plus 0.5em
  minus 0.4em\relax New York: Birkh{\"{a}}user Boston, 1990.

\bibitem{toi}
I.~S. Gradshteyn and I.~M. Ryzhik, \emph{Table of integrals, series and
  products}.\hskip 1em plus 0.5em minus 0.4em\relax New York: Academic Press,
  1966.

\bibitem{aofrvs}
M.~D. Springer, \emph{Algebra of random variables}.\hskip 1em plus 0.5em minus
  0.4em\relax New York: Wiley, 1979.

\end{thebibliography}
\end{document}